\documentclass[3p,review]{elsarticle}

\usepackage{lineno,hyperref}
\usepackage{graphicx,amssymb,arydshln,subeqnarray,subfigure}
\usepackage{url}
\usepackage{amsbsy,amsmath,amsfonts,cases,bm,caption,color}
\usepackage[ruled]{algorithm2e} 
\modulolinenumbers[5]

\journal{Journal of the Franklin Institute}

\begin{document}

\begin{frontmatter}

\title{Robust time-of-arrival localization via ADMM}

\author[mymainaddress]{Wenxin~Xiong\corref{mycorrespondingauthor}}
\cortext[mycorrespondingauthor]{Corresponding author}
\ead{w.x.xiong@outlook.com}

\author[mymainaddress]{Christian~Schindelhauer}
\ead{schindel@informatik.uni-freiburg.de}

\author[mysecondaryaddress]{Hing~Cheung~So\fnref{fn3}}
\ead{hcso@ee.cityu.edu.hk}

\address[mymainaddress]{Department of Computer Science, University of Freiburg, Freiburg 79110, Germany}
\address[mysecondaryaddress]{Department of Electrical Engineering, City University of Hong Kong, Hong Kong, China}

\fntext[fn3]{EURASIP Member.}

\begin{abstract}
This article considers the problem of source localization (SL) using possibly unreliable time-of-arrival (TOA) based range measurements. Adopting the strategy of statistical robustification, we formulate TOA SL as minimization of a versatile loss that possesses resistance against the occurrence of outliers. We then present an alternating direction method of multipliers (ADMM) to tackle the nonconvex optimization problem in a computationally attractive iterative manner. Moreover, we prove that the solution obtained by the proposed ADMM will correspond to a Karush-Kuhn-Tucker point of the formulation when the algorithm converges, and discuss reasonable assumptions about the robust loss function under which the approach can be theoretically guaranteed to be convergent. Numerical investigations demonstrate the superiority of our method over many existing TOA SL schemes in terms of positioning accuracy and computational simplicity. In particular, the proposed ADMM achieves estimation results with mean square error performance closer to the Cram\'{e}r-Rao lower bound than its competitors in our simulations of impulsive noise environments.
\end{abstract}


\end{frontmatter}

%\linenumbers

\section{Introduction}

Source localization (SL) refers to estimating the position of a signal-emitting target from measurements collected using multiple spatially separated sensors \cite{HCSo2}. Owing to its great significance to a lot of location-based applications, such as emergency assistance \cite{JHReed}, asset tracking \cite{FHoeflinger}, Internet of Things \cite{SLi}, and radar \cite{PSetlur}, the SL problem has received much attention in the literature over the past decades \cite{FZafari}. Depending on the measurements being used, methods for SL can be roughly divided into two groups: range- and direction-based. The category of range-based SL techniques, as the term \textit{range} suggests, involves the use of a variety of distance-related location data, including time-of-arrival (TOA) \cite{KWCheung,FKWChan,ABeck2,KWCheung2,KWKLui,JLiang2,YMPun,WHFoy,AColuccia,WXiongCVX}, time-difference-of-arrival \cite{ABeck2,JLiang2,WHFoy,DJTorrieri,MRGholami,GWang2,WXiong3}, time-sum-of-arrival \cite{JLiang2,MEinemo,RAmiri1,ZShi2,WXiongGRSL,WXiongTGRS,WXiongBP,WXiongBP2}, and received signal strength (RSS) \cite{AColuccia,WXiongCVX,WXiongIoT}, to determine the source position. Among these schemes, the time-based ones exploit the signal propagation time, while their RSS counterparts harness the signal power strength to acquire distance-type sensor observations. This dichotomy allows time-based SL solutions to excel in precision at the cost of more stringent synchronization requirements, while RSS-based approaches prioritize simplicity and inexpensiveness in implementation. In contrast, direction-based SL techniques utilize the source's bearings relative to the sensors to fulfill the positioning task \cite{DJTorrieri,AGabbrielli}. They, like methods built upon the RSS measurement model, can dispense with clock synchronization. However, direction-based positioning mandates the installation of an antenna array at each sensor for angle-of-arrival estimation, resulting in a somewhat higher hardware cost compared to its range-based rivals.

TOA-based SL takes center stage in this contribution, given its stature as one of the most widely adopted and intensively examined distance-based positioning schemes. Such a choice is further justified by the widespread availability of smartphones as well as other mobile devices that possess communication and information processing capabilities, most of which can obtain TOAs without encountering significant technical challenges \cite{FZafari}. Under the Gaussian noise assumption, least squares (LS) has undoubtedly been the very thing one needs to deliver statistically meaningful estimation results. In this vein, a wealth of LS techniques have been proposed in the literature, including algebraic explicit and exact solutions \cite{KWCheung,FKWChan,ABeck2}, convex programming approaches \cite{KWCheung2,KWKLui}, and local optimization algorithms \cite{JLiang2,YMPun,WHFoy}, to name but a few.

Statistical model mismatch will appear in real-world scenarios where the Gaussian noise assumption may be violated due to the presence of unreliable sensor data \cite{IGuvenc}, greatly deteriorating the performance of the LS methodology \cite{AMZoubir}. To reduce the negative impact of non-Gaussian measurement errors on positioning accuracy, researchers focusing on TOA-based SL have explored the strategies of joint estimation of location coordinates and a balancing parameter \cite{WXiongBP,WXiongBP2,STomic2,GWang}, worst-case LS criterion \cite{GWang,SZhang,STomic1}, and robust statistics \cite{WXiongCVX,WXiongIoT,CHPark3,WXiong,CSoares,HWang,WXiongMCC,WXiongImp}. Here, we are particularly interested in the statistical robustification of the TOA-based LS position estimator. Countermeasures of this type have been attracting increasing attention in recent years, mainly because of their relatively low prior knowledge requirements and low complexity of implementation.

Consider a single-source localization system deployed in $H$-dimensional space. It consists of an emitting source that is to be located, with unknown position $\bm{x} \in \mathbb{R}^{H}$, and $L$ sensors with signal receiving capabilities and known positions $\{ \bm{x}_{i} \in \mathbb{R}^{H} | i = 1,...,L \}$. Synchronization is guaranteed not only among the sensors themselves but also between the source and each sensor, so that the TOAs of the emitted signal at all $L$ sensors can be estimated. Specifically, the TOA-based range measurements are modeled as \cite{HCSo2}
\begin{equation}{\label{ch:ADMM:eqmdl_TOA}}
	r_{i} = {\| \bm{x} - \bm{x}_{i} \|}_2 + e_{i},~i = 1,...,L,
\end{equation}
where ${\| \cdot \|}_{2}$ denotes the $\ell_2$-norm and $e_{i}$ accounts for the observation uncertainty associated with the $i$th source-sensor path. Under the assumption that $\{ e_{i} \}$ are uncorrelated zero-mean Gaussian processes with variances $\{ \sigma_{i}^2 \}$ and in a maximum likelihood (ML) sense \cite{SMKay}, the TOA-based SL problem is mathematically stated as \cite{JLiang2}
\begin{equation}{\label{ch:ADMM:eqfm_ell_2}}
\min_{\bm{x}}~\sum_{i=1}^{L} \tfrac{(r_{i}-{\| \bm{x} - \bm{x}_{i} \|}_2)^2}{\sigma_{i}^2}.
\end{equation}

As mentioned earlier, the $\ell_2$-norm based estimator (\ref{ch:ADMM:eqfm_ell_2}) is vulnerable to outlying (unreliable) data, which manifest themselves as range measurements immersed in non-Gaussian, potentially bias-like errors. Common causes of them in the localization context include non-line-of-sight and multipath propagation of signals, interference, sensor malfunction, and malicious attacks \cite{PSetlur,IGuvenc,AMZoubir,TWang}. To achieve resistance against the occurrence of outliers in such adverse situations, (\ref{ch:ADMM:eqfm_ell_2}) was statistically robustified in \cite{WXiongCVX,WXiongIoT,CSoares,HWang,WXiongMCC,WXiongImp} by substituting the $\ell_2$ loss $(\cdot)^2$ with a certain fitting error measure less sensitive to biased $\{ r_{i} \}$, i.e.,
\begin{equation}{\label{ch:ADMM:eqfm_vers_loss}}
	\min_{\bm{x}}~\sum_{i=1}^{L}f(r_{i}-{\| \bm{x} - \bm{x}_{i} \|}_2),
\end{equation}
where frequently utilized options for $f(\cdot)$ encompass the $\ell_1$ loss $|\cdot|$ \cite{WXiongCVX,HWang}, the Huber loss:
\begin{equation}{\label{ch:ADMM:eqHuber_loss}}
	f_{R_{i}}(\cdot) = \left\{ \begin{aligned} &(\cdot)^2,~&\textup{for}~|\cdot| \leq R_i\\ &2R_{i}|\cdot|-R_i^2,~&\textup{for}~|\cdot| > R_i \end{aligned} \right.
\end{equation}
with radius $R_{i} > 0$ \cite{WXiongCVX,CSoares}, the Welsch loss $f_{\textup{W}}(\cdot) = \exp \left( - \tfrac{(\cdot)^2}{2\sigma_{\textup{W}}^2} \right)$ with parameter $\sigma_{\textup{W}}$ \cite{WXiongMCC}, and the $\ell_p$ loss ${|\cdot|}^{p}$ for $1 \leq p<2$ \cite{WXiongIoT,WXiongImp}. Below, let us provide a concise overview of several well-established optimization solutions available in the literature for addressing (\ref{ch:ADMM:eqfm_vers_loss}). They serve as competitors of our proposal, as we will discuss in Sections \ref{ch:ADMM:SCCA} and \ref{ch:ADMM:SR}. By forming the associated epigraph problem \cite{SBoyd1}:
\begin{align}{\label{ch:ADMM:eqform_l1_epi_ref}}
\min_{\bm{x},\bm{\upsilon}=[\upsilon_{1},...,\upsilon_{L}]^T \in \mathbb{R}^{L}}~\underbrace{\sum_{i=1}^{L}\upsilon_{i}}_{\textup{affine}},~~\textup{s.t.}~&\underbrace{r_{i} - \upsilon_{i}}_{\textup{affine}} - \underbrace{{\| \bm{x} - \bm{x}_i \|}_{2}}_{\textup{convex}} \leq 0,~i = 1,...,L,\nonumber\\
&\underbrace{{\| \bm{x} - \bm{x}_i \|}_{2}}_{\textup{convex}} - \underbrace{r_{i} - \upsilon_{i}}_{\textup{affine}} \leq 0,~i = 1,...,L,
\end{align}
one may directly apply second-order cone programming (SOCP), a classic convex approximation technique, or difference-of-convex programming (DCP) \cite{WXiongCVX} to address the $\ell_1$-minimization version of (\ref{ch:ADMM:eqfm_vers_loss}):
\begin{equation}{\label{ch:ADMM:eqfm_ell_1}}
	\min_{\bm{x}}~\sum_{i=1}^{L}|r_{i}-{\| \bm{x} - \bm{x}_{i} \|}_2|.
\end{equation}
The epigraph form (\ref{ch:ADMM:eqform_l1_epi_ref}) plays a crucial role as the basis for deriving (\ref{ch:ADMM:eqform_l1_epi_SOCPconv}) and (\ref{ch:ADMM:eqsubp_SOCP}), which represent the SOCP-based solution and DCP-based solution for (\ref{ch:ADMM:eqfm_ell_1}), respectively \cite{WXiongCVX}. Smoothed approximations through the composition of natural logarithm and hyperbolic cosine functions \cite{HWang}, on the other hand, render the Lagrange-type neurodynamic optimization framework of Lagrange programming neural network (LPNN) \cite{SZhangLPNN} applicable to
\begin{equation}{\label{ch:ADMM:eqfm_ell_1_smoothed}}
	\min_{\bm{x}}~\sum_{i=1}^{L}f_1 (r_{i}-{\| \bm{x} - \bm{x}_{i} \|}_2),
\end{equation}
where
\begin{equation}
	f_1 (z) = \tfrac{\log \left( \left( \exp{(\nu z)} + \exp{(-\nu z)} \right) / 2 \right)}{\nu}
\end{equation}
with parameter $\nu > 0$ is also known as (a.k.a.) the smoothed $\ell_1$ loss \cite{LZhao}. The idea of turning instead to the Huber convex underestimator for (\ref{ch:ADMM:eqmdl_TOA}) based on the composite loss $f_{R_{i}}([-(\cdot)]^{+})$ was conceptualized in \cite{CSoares}, where $[\cdot]^{+} = \max(\cdot,0)$ is the ramp function. Nonetheless, the emphasis of \cite{CSoares} is more on the extension of
\begin{equation}{\label{ch:ADMM:eqfm_HuberUE}}
	\min_{\bm{x}}\sum_{i=1}^{L}f_{R_{i}}([{\| \bm{x} - \bm{x}_{i} \|}_2 - r_{i}]^{+})
\end{equation}
to sensor network node localization, rather than single-source positioning. In \cite{WXiongMCC}, the challenging range-based Welsch loss minimization problem: 
\begin{equation}{\label{ch:ADMM:eqfm_RMCC}}
	\min_{\bm{x}}\sum_{i=1}^{L}f_{\textup{W}} \big( r_{i}-{\| \bm{x} - \bm{x}_{i} \|}_2 \big)
\end{equation}
was approximated by
\begin{equation}{\label{ch:ADMM:eqfm_SRMCC}}
	\min_{\bm{x}}\sum_{i=1}^{L}f_{\textup{W}} \big( r_{i}^2-{\| \bm{x} - \bm{x}_{i} \|}_2^2 \big)
\end{equation}
using the squared ranges, after which a half-quadratic (HQ) optimization algorithm was devised. The authors of \cite{WXiongImp} converted the $\ell_p$-minimization problem:
\begin{equation}{\label{ch:ADMM:eqfm_vers_loss_lp}}
	\min_{\bm{x}}~\sum_{i=1}^{L}{|r_{i}-{\| \bm{x} - \bm{x}_{i} \|}_2|}^p
\end{equation}
into an iteratively reweighted LS (IRLS) framework and applied sum-product message passing (MP) to solve the $\bm{x}$-update subproblem within each IRLS iteration in closed form.

While scaling poorly with the problem dimension is a well-known limitation of convex programming solutions whose implementation relies on interior-point methods, numerical realization of the neurodynamic approach in \cite{HWang} typically takes several thousands of iterations to reach equilibrium \cite{JLiang}. Despite the low-complexity advantage of the HQ algorithm in \cite{WXiongMCC}, the squared-range formulation (\ref{ch:ADMM:eqfm_SRMCC}) it adopts inherently suffers from impaired statistical efficiency. The MP-assisted IRLS \cite{WXiongImp} is computationally attractive, yet the lack of a theoretical foundation to support convergence of it may affect the soundness of the technique. With these concerns in mind, we are motivated to explore new avenues for coping with (\ref{ch:ADMM:eqfm_vers_loss}), especially in a way less computationally demanding but more statistically efficient and theoretically complete compared to the existing solutions.

In this contribution, we employ the alternating direction method of multipliers (ADMM) as our algorithmic strategy. ADMM is an iterative optimization solver which works by breaking down the original problem into several more easily solvable subproblems and, technically, it can be seen as an effort to combine the advantages of dual ascent and augmented Lagrangian methods \cite{SBoyd2}. Having originated in the 1970s with roots dating back to the 1950s \cite{SBoyd2,DGabay}, this simple but powerful algorithm has witnessed extensive use in the past few decades across various statistical and machine learning problems, including face recognition \cite{JXie}, subspace segmentation \cite{HZhang1}, portfolio optimization \cite{ZLShi}, image classification \cite{JQian}, and low-rank matrix recovery \cite{WXiongTGRS,HZhang2,HZhang3}. The key reasoning behind the adoption of ADMM in diverse applications of recent interest is that, while initially crafted for convex optimization, numerous extensions of it to the nonconvex setting still allow for lightweight yet reliable implementation and comprehensive theoretical analysis.

Our main contribution in this work is to develop an ADMM approach for tackling the generalized robust TOA positioning problem (\ref{ch:ADMM:eqfm_vers_loss}). As the versatility of the cost function $f(\cdot)$ will only be embodied in one of the subproblems that amounts to computing the proximal operator of it, we are able to adapt $f(\cdot)$ to specific noise environments with ease. In an analytical manner, we prove that if the presented ADMM is convergent, the limit point to which it converges will satisfy the Karush-Kuhn-Tucker (KKT) conditions for the equivalent constrained reformulation of (\ref{ch:ADMM:eqfm_vers_loss}). We also show the possibility of insuring theoretical convergence of the ADMM, under several additional assumptions about $f(\cdot)$. Numerical examples are included to corroborate the positioning accuracy and computational simplicity superiorities of our solution over its competitors. It should be pointed out that compared with the ADMM in \cite{JLiang2}, our scheme holds independent significance in the following two aspects. First, the algorithm in \cite{JLiang2} deals with the non-outlier-resistant formulation (\ref{ch:ADMM:eqfm_ell_2}), whereas we aim at solving (\ref{ch:ADMM:eqfm_vers_loss}) with a robustness-conferring fitting error measure. Second, we provide analytical convergence results of the presented ADMM. The authors of \cite{JLiang2}, in contrast, incorrectly assumed the nonconvex constrained optimization reformulation of (\ref{ch:ADMM:eqfm_ell_2}) to be convex, consequently omitting to probe into the convergence of their method under nonconvexity.

The rest of this contribution is organized as follows. The equivalent constrained reformulation of (\ref{ch:ADMM:eqfm_vers_loss}) and the ADMM solution to it are described in Section \ref{ch:ADMM:AD}. The complexity and convergence properties of the proposed ADMM are discussed in detail in Section \ref{ch:ADMM:SCCA}. Performance evaluations are conducted in Section \ref{ch:ADMM:SR}. Ultimately, Section \ref{ch:ADMM:C} concludes the article.

\section{Algorithm development}
\label{ch:ADMM:AD}

Let us begin by converting (\ref{ch:ADMM:eqfm_vers_loss}) into
\begin{align}{\label{ch:ADMM:eqell_p_ref}}
\min_{\bm{x},\bm{d}}\sum_{i=1}^{L}f(r_{i}-d_{i}),~\textup{s.t.}~{\left\| \bm{x} - \bm{x}_i \right\|}_2 = d_{i},~i = 1,...,L,
\end{align}
where $\bm{d} = \left[ d_1,...,d_L \right] \in \mathbb{R}^{L}$ is a dummy vector of decision variables for the source-sensor distances, independent of $\bm{x}$.

We further rewrite (\ref{ch:ADMM:eqell_p_ref}) as
\begin{subequations}{\label{ch:ADMM:eqell_p_ref_1}}
	\begin{align}
		\min_{\bm{x},\bm{d},\{ \bm{\beta}_{i} | {\| \bm{\beta}_{i} \|}_2^2 = 1 \}}& \sum_{i=1}^{L}f(r_{i}-d_{i}),\nonumber\\
		\textup{s.t.}~&\bm{x} - \bm{x}_i = \bm{\beta}_i \cdot d_{i},~i = 1,...,L,\label{ch:ADMM:eqell_p_ref_1_equa_cons}\\
		&d_{i} \geq 0,~i = 1,...,L,\label{ch:ADMM:eqell_p_ref_1_inequa_cons}
	\end{align}
\end{subequations}
by introducing an auxiliary vector $\bm{\beta} = \big[ \bm{\beta}_{1}^T,...,\bm{\beta}_{L}^T \big]^T \in \mathbb{R}^{HL}$ into the characterization of range information \cite{JLiang2}. Here, $\bm{\beta}_{i} \in \mathbb{R}^{H}$ is a unit vector that indicates the source's direction with respect to (w.r.t.) the $i$th sensor, a.k.a. the direction vector of arrival \cite{TKLe}.

The transformation from (\ref{ch:ADMM:eqell_p_ref}) to (\ref{ch:ADMM:eqell_p_ref_1}) bears some resemblance to the conversion from (\ref{ch:ADMM:eqell_p_ref}) to a formulation, where the Euclidean distances are re-expressed in a quadratic form. This conversion, in fact, is a widely adopted preprocessing step in the domain of range-based localization \cite{KWCheung2,ZShi2}, aimed at eliminating the cumbersome Euclidean norm terms in the initial formulation and simplifying the associated constrained optimization problem. In order to achieve an \textit{equivalent} reformulation during this simplification process, it is crucial to consider the quadratic form of the Euclidean distance constraints along with the nonnegativity constraints on the distance-representing variables \cite{KWCheung2,ZShi2}. In other words, it would be mathematically less rigorous to casually disregard these nonnegativity constraints without the backing of appropriate theoretical justifications, as demonstrated by the Appendix in \cite{KWCheung2} and Theorem 1 in \cite{ZShi2}. We therefore follow this tradition in our transformation from (\ref{ch:ADMM:eqell_p_ref}) to (\ref{ch:ADMM:eqell_p_ref_1}).

For (\ref{ch:ADMM:eqell_p_ref_1}), we construct the following augmented Lagrangian with constraints:
\begin{align}{\label{ch:ADMM:eqAL}}
&\mathcal{L}_\rho \left(\bm{x}, \bm{d},\bm{\beta}, \bm{\lambda} \right)\!=\!\sum_{i=1}^{L}f(r_{i} - d_{i})\!+\!\sum_{i = 1}^{L} \bm{\lambda}_i^T (\bm{x} - \bm{x}_i\!-\!\bm{\beta}_i \cdot d_{i}) +  \tfrac{\rho}{2} \sum_{i = 1}^{L} {\big\| \bm{x} - \bm{x}_i - \bm{\beta}_i \cdot d_{i} \big\|}_2^2,\nonumber\\
&\textup{s.t.}~{\| \bm{\beta}_{i} \|}_2^2 = 1,~i = 1,...,L,~~\textup{(\ref{ch:ADMM:eqell_p_ref_1_inequa_cons})},
\end{align}
where $\bm{\lambda} = \left[ \bm{\lambda}_1^T,...,\bm{\lambda}_L^T \right]^T \in \mathbb{R}^{HL}$ is a vector containing the Lagrange multipliers for (\ref{ch:ADMM:eqell_p_ref_1_equa_cons}) and $\rho > 0$ the augmented Lagrangian parameter. Splitting the primal variables into two parts, the ADMM for solving (\ref{ch:ADMM:eqell_p_ref_1}) consists of the following iterative steps:
\begin{subequations}{\label{ch:ADMM:eqcase}}
	\begin{align}
	&{\bm{x}}^{(k+1)} = \arg\min_{\bm{x}} \mathcal{L}_\rho \big(\bm{x}, {\bm{d}}^{(k)}, \bm{\beta}^{(k)}, \bm{\lambda}^{(k)} \big), \label{ch:ADMM:eqpx} \\
	&\big( {\bm{d}}^{(k+1)}, \bm{\beta}^{(k+1)} \big) = \arg\min_{{\bm{d}}, \bm{\beta}} \mathcal{L}_\rho \big(\bm{x}^{(k+1)}, {\bm{d}}, \bm{\beta}, \bm{\lambda}^{(k)} \big),~~\textup{s.t.}~\textup{(\ref{ch:ADMM:eqell_p_ref_1_inequa_cons})},~{\| \bm{\beta}_{i} \|}_2^2 = 1,~i = 1,...,L, \label{ch:ADMM:eqpdbeta} \\
	&\bm{\lambda}_i^{(k+1)} = \rho \big( \bm{x}^{(k+1)} - \bm{x}_i - \bm{\beta}_i^{(k+1)} \cdot d_i^{(k+1)} \big) + \bm{\lambda}_i^{(k)},~i = 1,...,L,\label{ch:ADMM:eqdlambda}
	\end{align}
\end{subequations}
where the iteration index is indicated by $(\cdot)^{(k)}$. To be specific, (\ref{ch:ADMM:eqpx}) and (\ref{ch:ADMM:eqpdbeta}) sequentially minimize the augmented Lagrangian (\ref{ch:ADMM:eqAL}) w.r.t. the primal variables, after which (\ref{ch:ADMM:eqdlambda}) updates the dual variables via a gradient ascent with step size $\rho$. We note that there are only two (namely, $\bm{x}$ and $( \bm{d}, \bm{\beta} )$) rather than three primal blocks in the ADMM governed by (\ref{ch:ADMM:eqcase}). By comparison, natural extension of the basic two-block ADMM to multi-block cases may result in divergence \cite{CChen}.

More detailed explanations for (\ref{ch:ADMM:eqpx}) and (\ref{ch:ADMM:eqpdbeta}) are given as follows.

By ignoring the constant terms independent of $\bm{x}$, the subproblem (\ref{ch:ADMM:eqpx}) can be simplified into an LS form
\begin{align}{\label{ch:ADMM:eqsubp1_ref}}
	\min_{\bm{x}}&~\tfrac{\rho}{2} \sum_{i = 1}^{L} {\big\| \bm{x} - \bm{w}_i^{(k)}\big\|}_2^2,
\end{align}
where 
\begin{equation}
	\bm{w}_i^{(k)} = \bm{x}_i + \bm{\beta}_i^{(k)} \cdot d_i^{(k)} -  \tfrac{1}{\rho} \bm{\lambda}_i^{(k)},~i = 1,...,L.
\end{equation}
It has the following closed-form solution:
\begin{equation}
	{\bm{x}}^{(k+1)} = \sum_{i=1}^{L} \bm{w}_i^{(k)} / L.
\end{equation}

Similarly, another subproblem (\ref{ch:ADMM:eqpdbeta}) is re-expressed as
\begin{align}{\label{ch:ADMM:eqsubp2_ref}}
	\min_{\bm{d}, \bm{\beta}}~\sum_{i = 1}^{L} f(r_{i} - d_{i}) + \sum_{i = 1}^{L}  \tfrac{\rho}{2} {\big\| \bm{v}_i^{(k+1)} - \bm{\beta}_i \cdot d_{i} \big\|}_2^2,~~\textup{s.t.}~{\| \bm{\beta}_{i} \|}_2^2 = 1,~i = 1,...,L,~\textup{(\ref{ch:ADMM:eqell_p_ref_1_inequa_cons})},
\end{align}
where
\begin{equation}
	\bm{v}_i^{(k+1)} = \bm{x}^{(k+1)} - \bm{x}_i +  \tfrac{1}{\rho} \bm{\lambda}_i^{(k)}.
\end{equation}
Exploiting the particular structure of (\ref{ch:ADMM:eqsubp2_ref}), the optimal $\bm{\beta}$ can be obtained first as
\begin{align}{\label{ch:ADMM:eqUpd_UV2}}
	\bm{\beta}_i^{(k+1)} = {\bm{v}_i^{(k+1)}}\big/{\big\| \bm{v}_i^{(k+1)} \big\|_2},~i = 1,...,L.
\end{align}
Thus, (\ref{ch:ADMM:eqsubp2_ref}) is reduced to
\begin{align}{\label{ch:ADMM:eqsubp2_rd}}
	\min_{\bm{d}}~\sum_{i = 1}^{L} f(r_{i} - d_{i}) + \sum_{i = 1}^{L}  \tfrac{\rho}{2} {\big( \big\| \bm{v}_i^{(k+1)} \big\|_2 - d_{i} \big)}^2,~~\textup{s.t.}~\textup{(\ref{ch:ADMM:eqell_p_ref_1_inequa_cons})},
\end{align}
which is separable w.r.t. the partition of $\bm{d}$ into its $L$ elements and leads to the following $L$ subproblems:
\begin{align}{\label{ch:ADMM:eqsubp2_rd_ref}}
	\min_{d_{i} \geq 0} f(r_{i} - d_{i})\!+\!\tfrac{\rho}{2} {\big( \big\| \bm{v}_i^{(k+1)} \big\|_2 - d_{i} \big)}^2,~i\!=\!1,...,L.
\end{align}
With the help of the following proposition, we can simplify (\ref{ch:ADMM:eqsubp2_rd_ref}) to a certain degree to facilitate the calculations.

\textsc{Proposition 1.} Under mild assumptions about the robust loss and range observations that $f(\cdot)$ is an even function strictly increasing on the nonnegative semi-axis and $\{ r_{i} \}$ are always nonnegative, coping with (\ref{ch:ADMM:eqsubp2_rd_ref}) will be equivalent to handling
\begin{align}{\label{ch:ADMM:eqsubp2_rd_ref_noncons}}
	\min_{d_{i}} f(r_{i} - d_{i})\!+\!\tfrac{\rho}{2} {\big( \big\| \bm{v}_i^{(k+1)} \big\|_2 - d_{i} \big)}^2,~i\!=\!1,...,L.
\end{align}

\textsc{Proof.} See \ref{Sect_AppenA}.

It is worth noting that the assumptions made in Proposition 1 are actually rather common (see our justification in the following). In such cases (where Proposition 1 applies), solving (\ref{ch:ADMM:eqsubp2_rd_ref}) boils down to computing the proximal operator of $f(\cdot)$, namely,
\begin{equation}{\label{ch:ADMM:eqproxlptp}}
	d_{i}^{(k+1)} = r_{i} - \mathcal{P}_{(1/\rho) f(\cdot)} \big( r_{i} - \big\| \bm{v}_i^{(k+1)} \big\|_2 \big),
\end{equation}
complying with the definition of proximal mapping below.

\textsc{Definition 1.} The proximal mapping of a function $f: \mathbb{R} \rightarrow \mathbb{R}$ with parameter $\tau > 0$ for any $b \in \mathbb{R}$ is \cite{NParikh}
\begin{equation}{\label{ch:ADMM:eqprox_gen}}
	\mathcal{P}_{\tau f(\cdot)} (b) = \arg \min_{a \in \mathbb{R}} f(a) +  \tfrac{1}{2\tau} (a - b)^2.
\end{equation}

\begin{table*}[!t]
	\renewcommand{\arraystretch}{1}
	\caption{Proximal computations for $\ell_p$ ($1 \leq p \leq 2$) and Huber loss functions}
	\label{ch:ADMM:table_CPC}
	\centering
	\begin{tabular}{|c|c|c|c|}
		\hline
		Loss type & Ref. & $f(\cdot)$ & $\mathcal{P}_{\tau f(\cdot)} (b)$ \\
		\hline
		$\ell_1$ & \cite{NParikh} & $|\cdot|$ & $\big[ b - \tau \big]^{+} - \big[ -b - \tau \big]^{+}$ \\
		\hline
		$\ell_p$ with $1 < p < 2$ & \cite{WJZeng} & ${|\cdot|}^{p}$ & $\left\{ \begin{aligned} &\arg \min_{a \in \{ 0, a^{+} \}} {|a|}^{p} +  \tfrac{1}{2\tau} (a - b)^2,~&\textup{if}~b \geq 0,\\
		&\arg \min_{a \in \{ a^{-}, 0 \}} {|a|}^{p} +  \tfrac{1}{2\tau} (a - b)^2,~&\textup{if}~b < 0. \end{aligned} \right.$ \\
		\hline
		$\ell_2$ & \cite{ABeck} & ${(\cdot)}^2$ & $\tfrac{b}{1+2\tau}$ \\
		\hline
		Huber & \cite{ABeck} & $f_{R_{i}}(\cdot)$ & $b -  \tfrac{2 \tau R_{i} b}{\max(|b|,R_{i} + 2 \tau R_{i})}$ \\
		\hline
	\end{tabular}
\end{table*}

Obviously, the computational simplicity of the proximal mapping procedure is crucial to the efficient update of $\bm{d}$ in each iteration of the proposed ADMM. In fact, the calculation of (\ref{ch:ADMM:eqprox_gen}) can be done in a relatively unencumbered manner or in closed form for many choices of $f(\cdot)$ exhibiting outlier-resistance \cite{NParikh,WJZeng,ABeck,PGong}. As summarized in Table \ref{ch:ADMM:table_CPC}\footnote{For the purpose of completeness, we also include the non-robust case of $p=2$, in which the $\ell_p$-minimization estimator will coincide with (\ref{ch:ADMM:eqfm_ell_2}), providing optimum estimation performance as long as the disturbances $\{ e_{i} \}$ are independent and identically distributed (i.i.d.) Gaussian processes.}, this article restricts the scope of discussion to two typical options: (i) the $\ell_p$ loss ${|\cdot|}^{p}$ for $1 \leq p < 2$ and (ii) the Huber function $f_{R_{i}}(\cdot)$. The two corresponding instantiations of (\ref{ch:ADMM:eqfm_vers_loss}) are then (\ref{ch:ADMM:eqfm_vers_loss_lp}) and
\begin{equation}{\label{ch:ADMM:eqfm_vers_loss_Huber}}
	\min_{\bm{x}}~\sum_{i=1}^{L}f_{R_{i}}(r_{i}-{\| \bm{x} - \bm{x}_{i} \|}_2),
\end{equation}
respectively. Our justification for them is given as follows. The $\ell_p$-minimization criterion with $1 \leq p < 2$ is known to show a considerable degree of robustness against outliers in a wide range of adverse environments \cite{WJZeng2,NHNguyen,WXiongImp,WJZeng}. Closely resembling its $\ell_1$ (resp. $\ell_2$) counterpart for large (resp. small) fitting errors, the Huber loss minimization scheme offers another means of allowing somewhat controllability therebetween \cite{DDeMenezes}. Most importantly, all these cost functions are compatible with our assumptions in Proposition 1.

We note that $a^{+}$ and $a^{-}$ in Table \ref{ch:ADMM:table_CPC} are the unique positive root of $\tfrac{a - b}{\tau} + pa^{p-1} = 0$ for $a \in [0, b]$ (when $b \geq 0$) and unique negative root of $\tfrac{a - b}{\tau} - p(-a)^{p-1} = 0$ for $a \in [b, 0]$ (when $b<0$), respectively, both of which can be conveniently found using simple root-finding algorithms (e.g., bisection or secant, among others). While these root-finding algorithms typically have only a constant time complexity of $\mathcal{O}(1)$ \cite{WJZeng}, the computational efficiency of the overall methodology may not always be optimal (e.g., they will lead to nested iterations in our $\ell_{p}$-norm-based ADMM use case). This discrepancy is evident in our CPU time comparisons, where we found that the $\ell_{p}$-norm-based ADMM method consistently required more runtime than the Huber-loss-based one. However, it is noteworthy that this increase in computational complexity does not significantly hinder the practicality of the corresponding ADMM algorithm. ADMM adopting the $\ell_{p}$ loss for $1<p<2$, even with the increased complexity, remains far more computationally efficient than several existing convex approximation approaches. Furthermore, in most cases, the Huber variant of our ADMM algorithm, which does not involve any iterative root-finding steps thanks to the closed-form proximal operator for Huber loss, already delivers decent performance. This means that there is usually limited need for invoking the $\ell_{p}$-norm version of the proposed ADMM technique (with iterative root-finding requirements) in typical TOA SL scenarios.

\begin{algorithm}[t]
	\SetAlgoLined
	\caption{ADMM for Robust TOA Positioning.}
	\label{ch:ADMM:Algorithm1}
	\KwIn{$\{ \bm{x}_{i} \}$, $\{ r_i \}$, $\rho$, and $\delta$.}
	
	{
		
		\textbf{Initialize:} $\bm{x}^{(0)}$, $\bm{d}^{(0)}$, $\bm{\beta}^{(0)}$, and $\bm{\lambda}^{(0)}$ randomly and feasibly.
		
		\For{$k=0,1,\cdots$}{

			\For{$i=1,\cdots,L$}{

				$\bm{w}_i^{(k)} \leftarrow \bm{x}_i + \bm{\beta}_i^{(k)} \cdot d_i^{(k)} -  \tfrac{1}{\rho} \bm{\lambda}_i^{(k)}$;

			}
		
			$\bm{x}^{(k+1)} \leftarrow \sum_{i=1}^{L} \bm{w}_i^{(k)} / L$;

			\For{$i=1,\cdots,L$}{

				$\bm{v}_i^{(k+1)} \leftarrow \bm{x}^{(k+1)} - \bm{x}_i +  \tfrac{1}{\rho} \bm{\lambda}_i^{(k)}$;

				$\bm{\beta}_i^{(k+1)} \leftarrow {\bm{v}_i^{(k+1)}}\big/{\big\| \bm{v}_i^{(k+1)} \big\|_2}$;
				
				$d_{i}^{(k+1)} \leftarrow r_{i} - \mathcal{P}_{(1/\rho) f(\cdot)} \big( r_{i} - \big\| \bm{v}_i^{(k+1)} \big\|_2 \big)$;

			}

			\For{$i=1,\cdots,L$}{

				$\bm{\lambda}_i^{(k+1)} \leftarrow \rho \big( \bm{x}^{(k+1)} - \bm{x}_i - \bm{\beta}_i^{(k+1)} \cdot d_i^{(k+1)} \big) + \bm{\lambda}_i^{(k)}$;

			}

			\textbf{Stop} if $\sum_{i=1}^{L}{\big\| \bm{x}^{(k)} - \bm{x}_i - \bm{\beta}_i^{(k)} \cdot d_i^{(k)} \big\|}_2 < \delta$
		
		}
		
		$\tilde{\bm{x}} \leftarrow \bm{x}^{(k+1)}$;

	}
	\KwOut{Estimate of source location $\tilde{\bm{x}}$.}
\end{algorithm}

The steps of ADMM for dealing with the robust TOA SL problem are summarized in Algorithm \ref{ch:ADMM:Algorithm1}, where $\delta$ is a user-defined tolerance to terminate the iterations.

\section{Complexity and convergence properties}
\label{ch:ADMM:SCCA}
The complexity and convergence properties of the proposed ADMM are discussed in detail in this section.

From Algorithm \ref{ch:ADMM:Algorithm1}, it is not difficult to find that the complexity of the ADMM is dominated by that of the $\bm{d}$-update steps, in which proximal mapping calculations are involved. Therefore, the total complexity of Algorithm \ref{ch:ADMM:Algorithm1} will be $\mathcal{O}(N_{\textup{ADMM}}L)$ if the loss function taken from Table \ref{ch:ADMM:table_CPC} admits closed-form $\bm{d}$-updates and $\mathcal{O}(N_{\textup{ADMM}}KL)$ otherwise, where $N_{\textup{ADMM}}$ represents the iteration number of the ADMM and $K$ the number of searches involved in the root-finding process at each ADMM iteration. Empirically, a few steps of the bisection method applied in each $\bm{d}$-update and a few tens to hundreds of ADMM iterations will already be sufficient to yield an accurate source location estimate.

By comparison, based on interior-point methods, solving the SOCP problem derived from (\ref{ch:ADMM:eqform_l1_epi_ref}) \cite{WXiongCVX}:
\begin{align}{\label{ch:ADMM:eqform_l1_epi_SOCPconv}}
\min_{\bm{x},\bm{\upsilon},\bm{d}}~\sum_{i=1}^{L}\upsilon_{i},~~\textup{s.t.}~&\big| r_{i} - d_{i} \big| \leq \upsilon_{i},~i = 1,...,L,\nonumber\\
&{\| \bm{x} - \bm{x}_i \|}_2 \leq d_{i},~i = 1,...,L,
\end{align}
implementing $N_{\textup{CCCP}}$ concave-convex procedure (CCCP) iterations in tackling the DCP problem (\ref{ch:ADMM:eqform_l1_epi_ref}) \cite{WXiongCVX}:
\begin{align}{\label{ch:ADMM:eqsubp_SOCP}}
\big( \bm{x}^{(k+1)}, \bm{\upsilon}^{(k+1)} \big)=\arg\min_{\bm{x},\bm{\upsilon}}~\sum_{i}^{L}\upsilon_{i},~~\textup{s.t.}&~r_{i} - \upsilon_{i} - {\| \bm{x}^{(k)} - \bm{x}_i \|}_2 - \big[ \bm{\nabla}_{[\bm{x}^T,\bm{\upsilon}^T]^T} \big( {\| \bm{x}^{(k)} - \bm{x}_i \|}_2 \big) \big]^T \nonumber\\
&~\big( \big[ \bm{x}^T,\bm{\upsilon}^T \big]^T - \big[ \big( \bm{x}^{(k)} \big)^T, \big( \bm{\upsilon}^{(k)} \big)^T \big]^T \big) \leq 0,~i = 1,...,L,\nonumber\\
&{\| \bm{x} - \bm{x}_i \|}_{2} - r_{i} - \upsilon_{i}^{(k)} - \big[ \bm{\nabla}_{[\bm{x}^T,\bm{\upsilon}^T]^T} \big( \upsilon_{i}^{(k)} \big) \big]^T \nonumber\\
&~\big( \big[ \bm{x}^T,\bm{\upsilon}^T \big]^T - \big[ \big( \bm{x}^{(k)} \big)^T, \big( \bm{\upsilon}^{(k)} \big)^T \big]^T \big) \leq 0,~i = 1,...,L,\nonumber\\
&~k = 0,...,N_{\textup{CCCP}}-1,
\end{align}
and realizing the ramp function based Huber convex underestimator (\ref{ch:ADMM:eqfm_HuberUE}) \cite{CSoares} result in $\mathcal{O}(L^{3.5})$, $\mathcal{O}(N_{\textup{CCCP}}L^{3.5})$, and $\mathcal{O}(L^{3.5})$ complexity, respectively \cite{GWang2}. The LPNN \cite{HWang}, the HQ technique \cite{WXiongMCC}, and the MP-assisted IRLS algorithm \cite{WXiongImp,WXiongIoT} are three other representative TOA positioning schemes from the literature that adopt the strategy of statistical robustification as well. They lead to $\mathcal{O}(N_{\textup{LPNN}}L)$, $\mathcal{O}(N_{\textup{HQ}}KL)$, and $\mathcal{O}(N_{\textup{IRLS}}L)$ complexity, respectively, where $N_{\textup{LPNN}}$ is the number of iterations in the numerical implementation of LPNN, $N_{\textup{HQ}}$ the number of steps needed for the HQ algorithm to converge, and $N_{\textup{IRLS}}$ that of the IRLS iterations. $N_{\textup{HQ}}$ typically takes a value of several tens, as does $K$ and $N_{\textup{IRLS}}$, while $N_{\textup{LPNN}}$ is generally in the range of several thousands \cite{WXiongMCC,WXiongImp,JLiang}. It turns out that the proposed ADMM with loss functions permitting closed-form proximal computations, the HQ algorithm in \cite{WXiongMCC}, and the IRLS in \cite{WXiongImp} are most computationally efficient.

With an appropriately selected value of $\rho$, Algorithm \ref{ch:ADMM:Algorithm1} is observed to be always convergent in our simulations. However, many existing theoretical analyses of ADMM convergence in nonconvex scenarios, such as \cite{YWang2}, may not be directly applicable to our context of robust TOA positioning due to two primary reasons. First, the optimization paradigms (of which convergence analyses were carried out) in many of these established analytical works differ from the constrained reformulation of the robust TOA positioning problem we address in our research. For instance, in \cite{YWang2}, any constraints on the optimization variables that are not linear constraints are treated as indicator functions and incorporated into the objective function of the optimization framework. In our problem (\ref{ch:ADMM:eqell_p_ref_1}), nevertheless, we explicitly impose unit-length constraints on the direction-indicating variables and nonnegativity constraints on the distance-representing variables, and even the equality constraint functions involved with (\ref{ch:ADMM:eqell_p_ref_1_equa_cons}) are not affine as per their definitions. Such discrepancies, in turn, result in distinct variable-update strategies between the ADMM algorithms considered in the relevant existing studies and our research. We note that similar dilemmas have been faced by many practitioners, despite empirically sound ADMM use cases in their respective contributions \cite{JLiang2,WXiong,WJZeng,YWang}. In what follows, we will present our own analytic proofs for the convergence of Algorithm \ref{ch:ADMM:Algorithm1} in lieu of pinning all hopes on the generally applicable results from the literature.

Establishing the equivalence between (\ref{ch:ADMM:eqell_p_ref_1}) and
\begin{align}{\label{ch:ADMM:eqell_p_ref_2}}
\min_{\bm{x},\bm{d},\{ \bm{\beta}_{i} | {\| \bm{\beta}_{i} \|}_2^2 = 1 \}} \sum_{i=1}^{L}f(r_{i} - d_{i}),~~\textup{s.t.}~\bm{x} - \bm{x}_i = \bm{\beta}_i \cdot d_{i},~i = 1,...,L,
\end{align}
the following proposition points out that (\ref{ch:ADMM:eqell_p_ref_1_inequa_cons}) can be disregarded and we may shift our focus to the simplification (\ref{ch:ADMM:eqell_p_ref_2}) instead in the convergence analysis:

\textsc{Proposition 2.} Under the same assumptions as in Proposition 1, the formulations (\ref{ch:ADMM:eqell_p_ref_1}) and (\ref{ch:ADMM:eqell_p_ref_2}) are equivalent to one another.

\textsc{Proof.} The proof of Proposition 2 is similar to that of Proposition 1. Please see \ref{Sect_AppenB} for the details.

Next, we derive the following theorem for the optimality of tuples produced by Algorithm \ref{ch:ADMM:Algorithm1}:

\textsc{Theorem 1.} Let $\big\{ \bm{x}^{(k)}, \bm{d}^{(k)}, \bm{\beta}^{(k)}, \bm{\lambda}^{(k)} \big\}_{k=1,...}$ be the tuples of primal and dual variables generated by Algorithm \ref{ch:ADMM:Algorithm1}. If
\begin{equation}{\label{ch:ADMM:eqlim_assump}}
	\lim_{k \rightarrow \infty} \big\{ \bm{x}^{(k)}, \bm{d}^{(k)}, \bm{\beta}^{(k)}, \bm{\lambda}^{(k)} \big\} = ( \bm{x}^{\star}, \bm{d}^{\star}, \bm{\beta}^{\star}, \bm{\lambda}^{\star} ),
\end{equation}
then the limit $( \bm{x}^{\star}, \bm{d}^{\star}, \bm{\beta}^{\star}, \bm{\lambda}^{\star} )$ will satisfy the KKT conditions (a.k.a. the first-order necessary conditions \cite{JNocedal}) for (\ref{ch:ADMM:eqell_p_ref_2})\footnote{In cases where $f(\cdot)$ is non-differentiable at the origin, the condition (\ref{ch:ADMM:eqKKTcon_Lag2}) should be substituted with the inclusion: $\bm{0}_{L} \in \hat{\partial}_{\bm{d}} (\bm{x}^{\star}, \bm{d}^{\star},\bm{\beta}^{\star}, \bm{\lambda}^{\star})$ \cite{RFeng}, where $\hat{\partial}_{\bm{d}}(\cdot)$ denotes the generalized gradient of a function at $\bm{d}$ that takes into consideration the subdifferential calculus \cite{FHClarke}. For simplicity's sake, we confine our analyses here to the special case with a differentiable $f(\cdot)$. Nevertheless, it is worth pointing out that with slight modifications, the results can be readily extended to the scenarios in the presence of non-differentiability.}:
\begin{subequations}
	\begin{align}
	&\bm{\nabla}_{\bm{x}} \hat{\mathcal{L}} (\bm{x}^{\star}, \bm{d}^{\star},\bm{\beta}^{\star}, \bm{\lambda}^{\star}) = \bm{0}_H,\label{ch:ADMM:eqKKTcon_Lag1}\\
	&\bm{\nabla}_{\bm{d}} \hat{\mathcal{L}} (\bm{x}^{\star}, \bm{d}^{\star},\bm{\beta}^{\star}, \bm{\lambda}^{\star}) = \bm{0}_L,\label{ch:ADMM:eqKKTcon_Lag2}\\
	&\bm{\nabla}_{\bm{\beta}} \hat{\mathcal{L}} (\bm{x}^{\star}, \bm{d}^{\star},\bm{\beta}^{\star}, \bm{\lambda}^{\star}) = \bm{0}_{HL},\label{ch:ADMM:eqKKTcon_Lag3}\\
	&\bm{x}^{\star} - \bm{x}_i = \bm{\beta}_i^{\star} \cdot d_i^{\star},~i = 1,...,L,\label{ch:ADMM:eqKKTcon_cons}\\
	&{\| \bm{\beta}_{i}^{\star} \|}_2^2 = 1,~i = 1,...,L,\label{ch:ADMM:eqKKTcon_beta}
	\end{align}
\end{subequations}
where
\begin{align}
	\hat{\mathcal{L}}(\bm{x}, \bm{d},\bm{\beta}, \bm{\lambda}) = \sum_{i=1}^{L}f(r_{i} - d_{i}) + \sum_{i = 1}^{L} \bm{\lambda}_i^T (\bm{x} - \bm{x}_i - \bm{\beta}_i \cdot d_{i})
\end{align}
is the associated Lagrangian of (\ref{ch:ADMM:eqell_p_ref_2}).

\textsc{Proof.} See \ref{Sect_AppenC}.

\textsc{Remark 1.} Theorem 1 does not guarantee the convergence of the proposed ADMM. Instead, it only reveals the optimality of solution when the algorithm converges. To further improve the analytical completeness, one may borrow the idea from \cite{YWang} to relax the formulation (\ref{ch:ADMM:eqell_p_ref_2}) somewhat by introducing additional auxiliary variables and an extra quadratic penalty, whereby rigorously proving the convergence of the generated sequence is made possible. This will, however, give rise to degradation of the estimator's performance \cite{YWang} (i.e., the incompleteness of analytical convergence results can be seen as an acceptable trade-off for higher estimation accuracy). For avoiding such performance-impairing approximations to the source localization formulation, one would have to limit the scope of choice of $f(\cdot)$, as elaborated in the following statements.

\textsc{Theorem 2.} If $f(r_{i} - d_{i})$ is convex w.r.t. $d_{i}, \forall i \in \{ 1, \ldots ,L \}$, and for some constant $\bar{M}$, the gradient of $f(r_{i} - d_{i})$ satisfies that $\lvert \lvert\nabla_{d_i}f(r_{i} - d_{i})\vert_{d_i = d_{i,1}} \rvert + \lvert \nabla_{d_i}f(r_{i} - d_{i})\vert_{d_i = d_{i,2}} \rvert \rvert \leq \bar{M} {\lvert  d_{i,1} - d_{i,2} \rvert }$, the sequence:
\begin{equation}
	\big\{ \mathcal{L}_\rho \big( \bm{x}^{(k)}, \bm{d}^{(k)}, \bm{\beta}^{(k)}, \bm{\lambda}^{(k)} \big) \big\}_{k=1,...}
\end{equation}
will be monotonically nonincreasing as the iterations (\ref{ch:ADMM:eqpx})--(\ref{ch:ADMM:eqdlambda}) proceed, for a sufficiently large $\rho$.

\textsc{Proof.} See \ref{Sect_AppenD}.

\textsc{Theorem 3.} With the assumptions made about $f(\cdot)$ in Proposition 1 and Theorem 2, and an additional assumption that the gradient of $f(\cdot)$ at $d_i$ is Lipschitz continuous with a constant $M_{1}$, $\mathcal{L}_\rho \big( \bm{x}^{(k)}, \bm{d}^{(k)}, \bm{\beta}^{(k)}, \bm{\lambda}^{(k)} \big)$ is bounded from below for sufficiently large $\rho$.

\textsc{Proof.} See \ref{Sect_AppenE}.

\textsc{Corollary 1.} $\big\{ \mathcal{L}_\rho \big( \bm{x}^{(k)}, \bm{d}^{(k)}, \bm{\beta}^{(k)}, \bm{\lambda}^{(k)} \big) \big\}_{k=1,...}$ is convergent under the above assumptions about $f(\cdot)$ and $\rho$.

\textsc{Proof.} This corollary is established via Theorems 2 and 3. $\square$

Equipped with these promising tools, we are finally enabled to present the following theorem, which asserts that Algorithm \ref{ch:ADMM:Algorithm1} can be theoretically guaranteed to converge with certain reasonable assumptions.

\textsc{Theorem 4.} The sequence $\big\{ \bm{x}^{(k)}, \bm{d}^{(k)}, \bm{\beta}^{(k)}, \bm{\lambda}^{(k)} \big\}_{k=1,...}$ is convergent, namely, (\ref{ch:ADMM:eqlim_assump}) holds, under the same circumstances as in Corollary 1.

\textsc{Proof.} See \ref{Sect_AppenF}.

\textsc{Corollary 2.} Let us adopt the setting of Corollary 1. The solution obtained by Algorithm \ref{ch:ADMM:Algorithm1} corresponds to a KKT point of the nonconvex constrained optimization problem (\ref{ch:ADMM:eqell_p_ref_2}).

\textsc{Proof.} This corollary is established via Theorems 1 and 4. $\square$

\section{Simulation results}
\label{ch:ADMM:SR}

\begin{table*}[!t]
	\renewcommand{\arraystretch}{1}
	\caption{Robust TOA positioning algorithms considered in comparisons}
	\label{ch:ADMM:table_RTOA}
	\centering
	\begin{tabular}{|c|c|c|c|}
		\hline
		Description & Ref. & Problem & Abbr. \\
		\hline
		SOCP for $\ell_1$-minimization & \cite{WXiongCVX} & (\ref{ch:ADMM:eqform_l1_epi_SOCPconv}) & $\ell_1$-\texttt{SOCP} \\
		\hline
		CCCP-based DCP for $\ell_1$-minimization & \cite{WXiongCVX} & (\ref{ch:ADMM:eqsubp_SOCP}) & $\ell_1$-\texttt{DCP} \\
		\hline
		Huber convex underestimator & \cite{CSoares} & (\ref{ch:ADMM:eqfm_HuberUE}) & \texttt{Huber}-\texttt{CUE} \\
		\hline
		LPNN for smoothed $\ell_1$-minimization & \cite{HWang} & (\ref{ch:ADMM:eqfm_ell_1_smoothed}) & $\ell_1$-\texttt{LPNN} \\
		\hline
		HQ algorithm for Welsch loss minimization & \cite{WXiongMCC} & (\ref{ch:ADMM:eqfm_SRMCC}) & \texttt{Welsch}-\texttt{HQ} \\
		\hline
		MP-assisted IRLS for $\ell_p$-minimization ($1 \leq p < 2$) & \cite{WXiongImp} & (\ref{ch:ADMM:eqfm_vers_loss_lp}) & $\ell_p$-\texttt{IRLS} \\
		\hline
		Proposed ADMM for $\ell_p$-minimization ($1 \leq p \leq 2$) & -- & (\ref{ch:ADMM:eqfm_vers_loss_lp}) & $\ell_p$-\texttt{ADMM} \\
		\hline
		Proposed ADMM for Huber loss minimization & -- & (\ref{ch:ADMM:eqfm_vers_loss_Huber}) & \texttt{Huber}-\texttt{ADMM} \\
		\hline
	\end{tabular}
\end{table*}

In this section, we carry out computer simulations to evaluate the performance of Algorithm \ref{ch:ADMM:Algorithm1}. Comparisons are made between the proposed ADMM and several existing TOA-based location estimators that are also built upon robust statistics. Table \ref{ch:ADMM:table_RTOA} gives a summary of these methods, and we note that the convex approximation techniques are all realized using the MATLAB CVX package \cite{MGrant}. To implement the LPNN \cite{HWang}, we invoke the MATLAB routine \texttt{ode15s}, a variable-step, variable-order solver relying on the formulas for numerical differentiation of orders 1 to 5 \cite{LFShampine}.

An SL system with $H = 2$ and $L = 8$ is considered. Unless otherwise mentioned, the locations of the source and sensors are randomly generated inside an origin-centered 20 m $\times$ 20 m square area, in each of the Monte Carlo (MC) runs. The positioning accuracy metric is the root-mean-square error (RMSE), defined as
\begin{equation}
	\textup{RMSE} = \sqrt{\sum_{j=1}^{N_{\textup{MC}}} \tfrac{{ \left\| \tilde{\bm{x}}^{\{ j \}} - {\bm{x}}^{\{ j \}} \right\| }_{2}^{2}}{N_{\textup{MC}}}},
\end{equation}
where $N_{\textup{MC}}$ denotes the total number of MC runs and is fixed at $3000$ here, and $\tilde{\bm{x}}^{\{ j \}}$ represents the estimate of the source position, ${\bm{x}}^{\{ j \}}$, in the $j$th MC run. User-specified parameters of the ADMM are set as $\delta = {10}^{-5}$ and $\rho = 5$\footnote{It is essential to emphasize that the optimal selection of $\rho$ depends on both the problem scale and the specific setup of the loss function, i.e., careful consideration is required to ensure its optimality. For instance, when all variables in (\ref{ch:ADMM:eqell_p_ref_1}) are increased by a factor of 10, the last term of the augmented Lagrangian (\ref{ch:ADMM:eqAL}) will increase by a factor of 100, while the first two terms will increase by a factor of only 10 when using, e.g., the $\ell_{1}$ loss function. To maintain a balance between the two parts of the augmented Lagrangian, it is necessary to decrease $\rho$ to some extent, optimizing the numerical performance of ADMM. Nonetheless, for the sake of simplicity, our simulations exclusively employ a fixed $\rho$-setting with $\rho = 5$, though this value may not be optimal in every scenario. We anticipate a thorough exploration of the impacts of $\rho$ to be a topic for future research, as explained in Section \ref{ch:ADMM:C}.}. The tunable parameter $\sigma_{\textup{W}}$ of the Welsch loss is adaptively chosen according to the Silverman's heuristic \cite{WLiu}, in the same way as shown in \cite{WXiongMCC}. The smoothing parameter $\nu$ is set to 0.1 to ensure feasibility of the \texttt{ode15s} solver.

\begin{figure}[!t]
	\centering
	\includegraphics[width=6.5in]{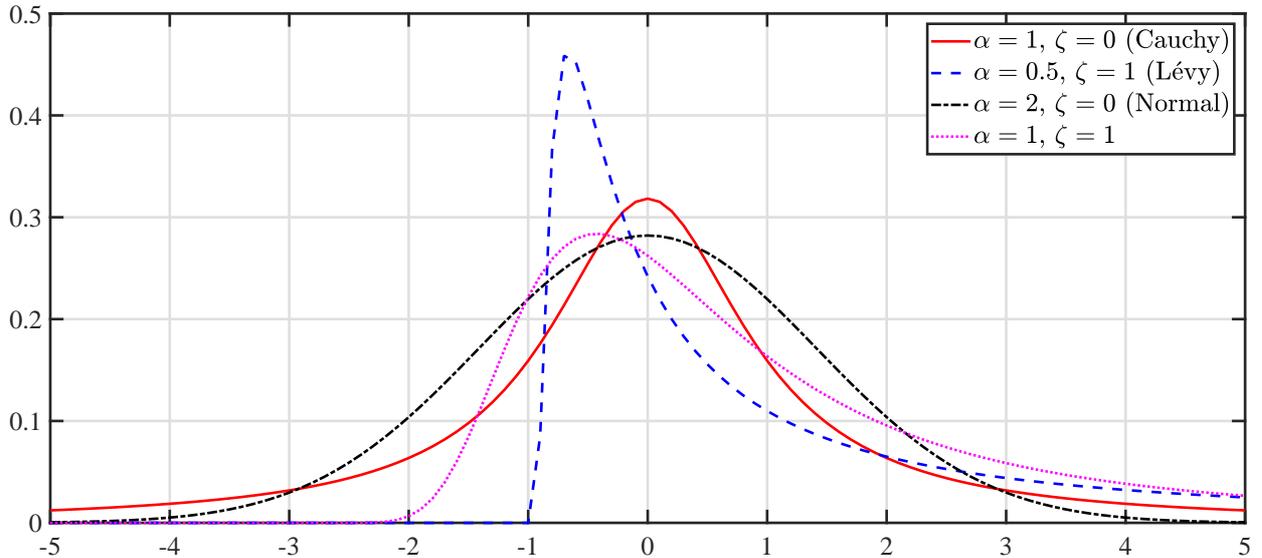}
	\caption{PDF plots for stable distributions with $\gamma = 1$ and $\mu = 0$.} 
	\label{Fig_PDFs}
\end{figure}

Taking into account the presence of outliers, we follow \cite{WXiongImp,NHNguyen} to use the class of $\alpha$-stable distributions for modeling $\{ e_{i} \}$ in (\ref{ch:ADMM:eqmdl_TOA}). Stable processes are well known for their suitability to characterize skewness and heavy-tailedness. Except for a few special instances, members of the stable distribution family do not have an explicit expression of probability density function (PDF). Instead, their PDF $p(z)$ is implicitly described through the inverse Fourier transform of the characteristic function $\Phi(t;\alpha,\zeta,\gamma,\mu)$:
\begin{align}
	p(z) =  \tfrac{1}{2\pi} \int_{-\infty}^{\infty}\Phi(t) \exp(-jzt)\textup{d}t,
\end{align}
where the detailed analytical parameterization of $\Phi(t)$ can be found in \cite{JPNolan}. As one may see, there are four parameters defining the family. The stability parameter, $0 < \alpha \leq 2$, controls the tails of the distribution. Generally speaking, the smaller the value of $\alpha$, the heavier the tails and the more impulsive the random variable being modeled. $\mu$ determines the location. The skewness parameter, $-1 \leq \zeta \leq 1$, is a measure of asymmetry. In the simplest case where $\zeta = 0$, the distribution becomes symmetric about its mean (which is, $\mu$, when $\alpha > 1$) and degenerates into the so-called symmetric $\alpha$-stable ($S \alpha S$) distribution. By contrast, the distribution is said to be right-skewed (resp. left-skewed) for $\zeta > 0$ (resp. $\zeta < 0$). $\gamma > 0$ is the scale parameter, measuring to what extent the distribution spreads out (similar to the variance of the normal distribution). For illustration purposes, Fig. \ref{Fig_PDFs} plots the PDF functions for several representative choices of $\alpha$ and $\zeta$. As in our case, the stable-distributed range measurement noise $\{ e_{i} \}$ is assumed to be i.i.d. with stability, skewness, scale, and location parameters $\alpha$, $\zeta = 0$, $\gamma$, and $\mu = 0$, respectively.

Because the variance of the stable distribution is undefined for $\alpha < 2$, we introduce the concept of generalized signal-to-noise ratio (GSNR) from \cite{WXiongImp} to quantify the relative noise level:
\begin{align}
	\textup{GSNR} = 10 \log_{10} \left(  \tfrac{\sum_{i=1}^{L} {\| \bm{x} - \bm{x}_i \|}_2^2}{L{\gamma}^{\alpha}} \right).
\end{align}
Furthermore, the square root of the trace of the MC-approximated Cram\'{e}r-Rao lower bound matrix (termed \texttt{RCRLB}) \cite{FYin1} is included in the comparison results to offer a benchmark for the accuracy of different robust position estimators in i.i.d. non-Gaussian noise. As it is in general difficult to work out a consistent schedule for adjusting the Huber radius under stably modeled non-Gaussianity, we simply follow \cite{YLiu} to assign a fixed value of 1 to $R_{i}$.

\begin{figure}[!t]
	\centering
	\includegraphics[width=6.5in]{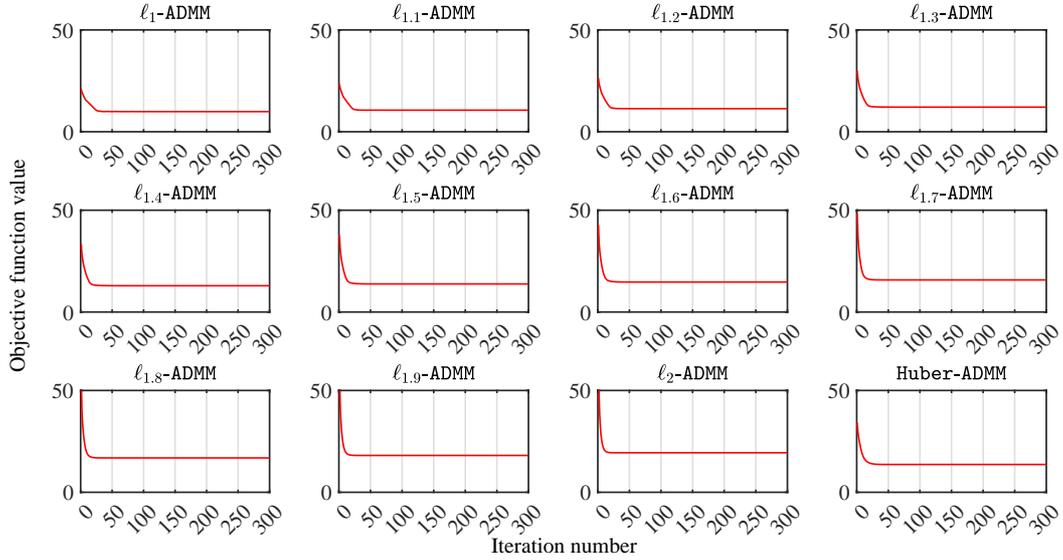}
	\caption{Objective function value versus iteration number.}
	\label{Fig_RMSE_Convergence1}
\end{figure}

\begin{figure}[!t]
	\centering
	\includegraphics[width=6.5in]{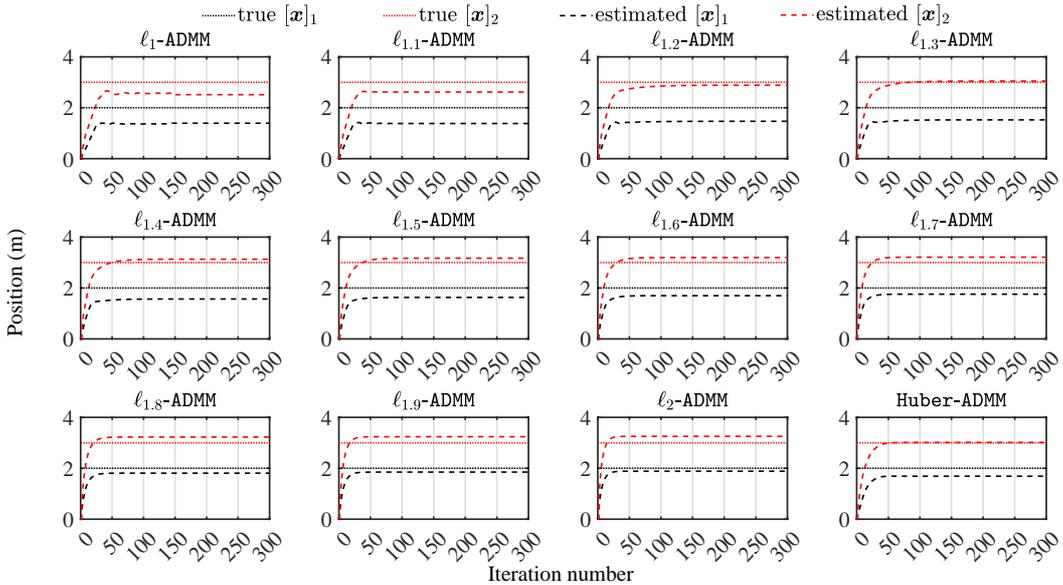}
	\caption{Position estimate versus iteration number.}
	\label{Fig_RMSE_Convergence2}
\end{figure}

Starting off, Figs. \ref{Fig_RMSE_Convergence1} and \ref{Fig_RMSE_Convergence2} show the convergence behavior of $\ell_p$-\texttt{ADMM} ($p$ taking different values from 1 to 2) and \texttt{Huber}-\texttt{ADMM} in a single MC run for $\alpha = 1.5$ and GSNR $=20$ dB. We should also note that for reproducibility, in this test $L=8$ sensors are evenly placed on the perimeter of the afore-defined square region and the source is deterministically deployed at $\bm{x} = [2,3]^T$ m. Both $\ell_{p}$-\texttt{ADMM} and \texttt{Huber}-\texttt{ADMM} are observed to rapidly decrease the objective function and converge to a point close to the true source position, within the first few tens of iterations. In our following simulation studies, the value of $p$ required by $\ell_{p}$-\texttt{ADMM} and $\ell_p$-\texttt{IRLS} will be set to the optimal one hinging on $\alpha$ \cite{YChen}, in a way analogous to \cite{WXiongImp,NHNguyen}.

\begin{figure}[!t]
	\centering
	\includegraphics[width=6.5in]{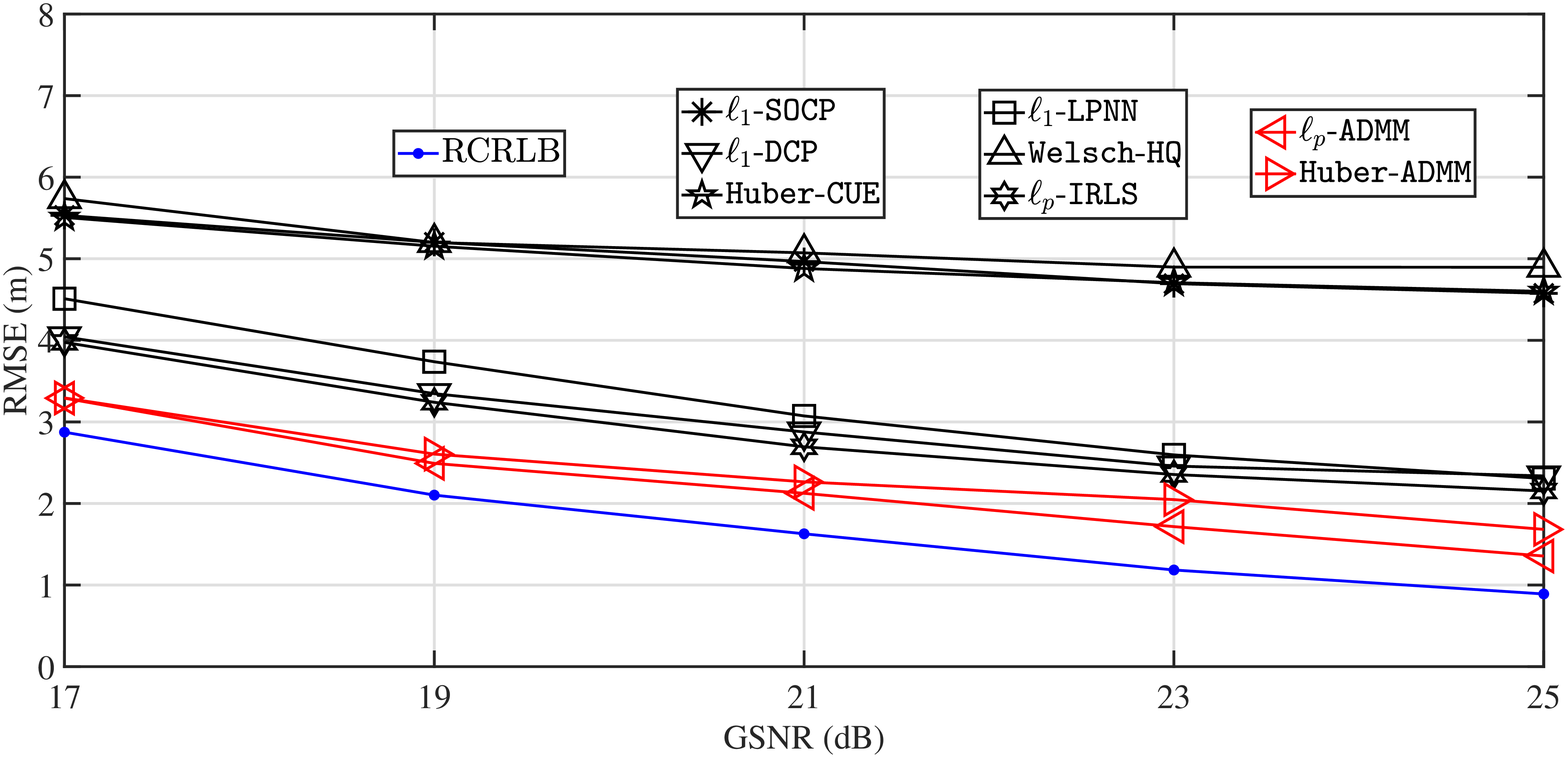}
	\caption{RMSE versus GSNR.} 
	\label{Fig_RMSE_GSNR}
\end{figure}

\begin{figure}[!t]
	\centering
	\includegraphics[width=6.5in]{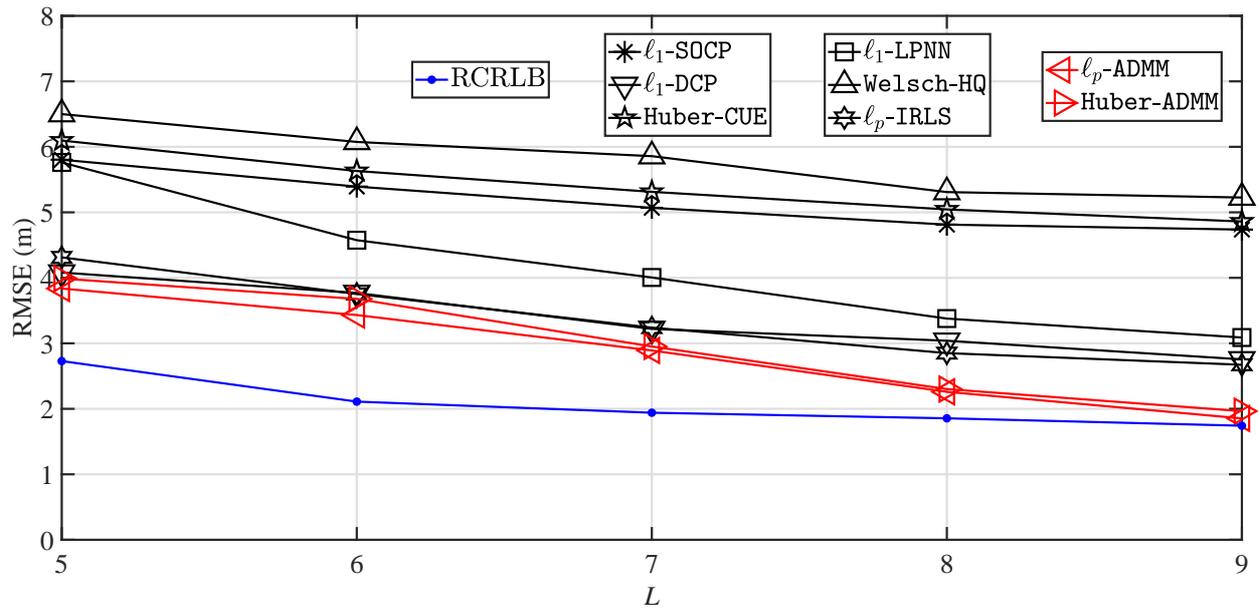}
	\caption{RMSE versus $L$.} 
	\label{Fig_RMSE_L}
\end{figure}

\begin{figure}[!t]
	\centering
	\includegraphics[width=6.5in]{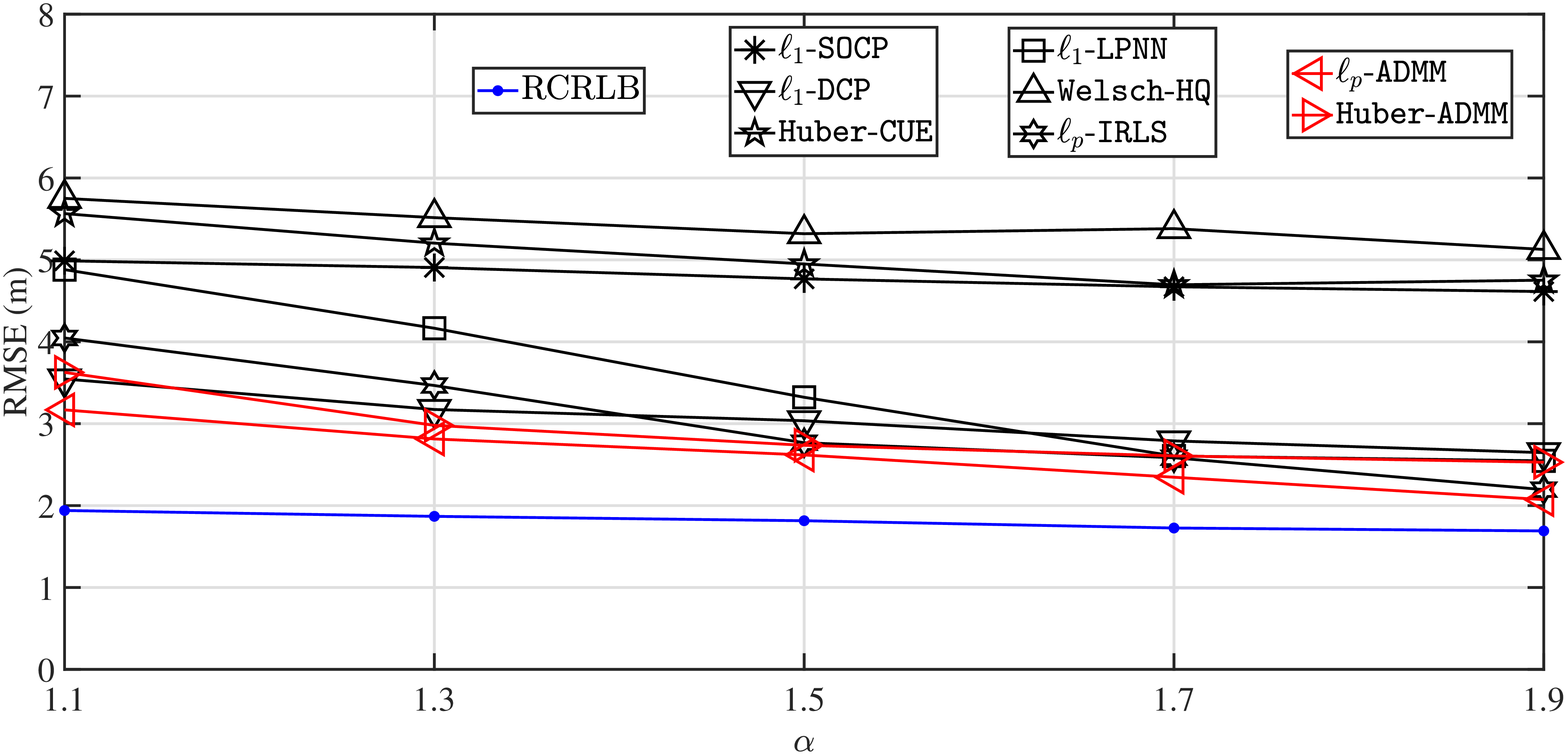}
	\caption{RMSE versus $\alpha$.} 
	\label{Fig_RMSE_alpha}
\end{figure}

Fig. \ref{Fig_RMSE_GSNR} depicts the RMSE versus GSNR $\in [17, 25]$ dB while fixing $\alpha$ at $1.5$. It is seen that the two proposed ADMM approaches deliver the lowest level of positioning errors for all GSNRs. Especially, the RMSE performance of $\ell_{p}$-\texttt{ADMM} is the closest to the \texttt{RCRLB} benchmark (with a small gap of about half a meter). Among the six competitors, $\ell_p$-\texttt{IRLS} produces the best results. This reconfirms the findings reported in \cite{WXiongImp}, that $\ell_p$-\texttt{IRLS} can be fairly statistically efficient in impulsive noise if $p$ is optly chosen. The direct nonlinear optimization based $\ell_1$-minimization techniques, $\ell_1$-\texttt{DCP} and $\ell_1$-\texttt{LPNN}, are slightly inferior to $\ell_p$-\texttt{IRLS} due to model mismatch but still capable of outperforming the remaining schemes, whose estimation performance looks relatively poorer. $\ell_1$-\texttt{LPNN} will not be comparable to $\ell_1$-\texttt{DCP} as the GSNR decreases from 25 dB, attributed to the smoothed approximations introduced in (\ref{ch:ADMM:eqfm_ell_1_smoothed}). Nonetheless, the former does not have tightness issues from which the traditional convex relaxation methods $\ell_1$-\texttt{SOCP} and \texttt{Huber}-\texttt{CUE} suffer, nor is it quite as statistically inefficient as \texttt{Welsch}-\texttt{HQ}.

Setting the GSNR to 20 dB and the impulsiveness-controlling parameter as $\alpha = 1.5$, we plot the RMSE versus the number of sensors used for localization, $L \in [5, 9]$, in Fig. \ref{Fig_RMSE_L}. Again, the performance of the two ADMM solutions is optimal. Although none of the estimators can attain \texttt{RCRLB}, our algorithms are still the closest ones to it, particularly in the scenarios with a large number of sensors. Fig. \ref{Fig_RMSE_alpha} further plots the RMSE as a function of $\alpha \in [1.1, 1.9]$ for $L = 8$ and GSNR $= 20$ dB. In this scenario, once more, $\ell_{p}$-\texttt{ADMM} stands out as superior compared to other approaches.

\begin{table}[!t]
	\renewcommand{\arraystretch}{1}
	\caption{Average CPU time for simulations conducted on a laptop with 16 GB of memory and a 4.7 GHz CPU}
	\label{table_CPU}
	\centering
	\begin{tabular}{|c|c|}
		\hline
		\textbf{Algorithm} & \textbf{Time} (s) \\
		\hline
		$\ell_1$-\texttt{SOCP} & 0.677 \\
		\hline
		$\ell_1$-\texttt{DCP} & 3.261 \\
		\hline
		\texttt{Huber}-\texttt{CUE} & 1.669 \\
		\hline
		$\ell_1$-\texttt{LPNN} & 0.158 \\
		\hline
		\texttt{Welsch}-\texttt{HQ} & 0.026 \\
		\hline
		$\ell_p$-\texttt{IRLS} & 0.003 \\
		\hline
		$\ell_p$-\texttt{ADMM} & 0.483 \\
		\hline
		\texttt{Huber}-\texttt{ADMM} & 0.010 \\
		\hline
	\end{tabular}
\end{table}

In Table \ref{table_CPU}, we list the per-sample average CPU time of the eight location estimators in the above simulations. It is seen that the running time can vary by several orders of magnitude. \texttt{Huber}-\texttt{ADMM} that allows for closed-form proximal mapping of $f(\cdot)$ and \texttt{Welsch}-\texttt{HQ} are the second and third fastest, respectively, taking only slightly more time than the state-of-the-art method $\ell_p$-\texttt{IRLS}. Because of the extra bisection procedure incorporated into the ADMM algorithm, $\ell_p$-\texttt{ADMM} may not be as computationally efficient as \texttt{Huber}-\texttt{ADMM} in actual CPU time comparisons. Nonetheless, its problem-solving process still takes less time than those of the traditional convex approximation techniques. In general, the simulation results agree with our complexity analysis in Section \ref{ch:ADMM:SCCA}.

\section{Conclusion and future research directions}
\label{ch:ADMM:C}
This article focused on the problem of outlier-resistant TOA SL. Based on the ADMM, we presented an iterative algorithm to handle the statistically robustified positioning formulation. Each iteration of our ADMM consists of two primal variable minimization steps and a dual variable update, all of which can be effortlessly implemented provided that the loss function relied on allows for convenient proximal computations. What is more, we proved that the ADMM will converge to a KKT point of the nonconvex constrained optimization problem under certain conditions, and verified that the LICQ holds at the point for nontrivial SL configurations. The superiority of the devised scheme over a number of robust statistical TOA SL methods in terms of positioning accuracy in impulsive noise and computational simplicity was demonstrated through simulations.

Future research endeavors could involve (i) deriving analytical convergence results under a milder setting of $\rho$, (ii) embarking on a more detailed exploration of the influences of $\rho$, and (iii) extending ADMM to the novel use case of robust direction-based SL with outliers.

\appendix

\section{Proof of Proposition 1}
\label{Sect_AppenA}
Denote the globally optimal solution to (\ref{ch:ADMM:eqsubp2_rd_ref_noncons}) by $\bm{d}^{*}$\footnote{We stipulate that the asterisk and star symbols in this paper apply to each element of the corresponding vectors by default.}. Straightforwardly, the proposition will hold if we are able to show that $d_{i}^{*} \geq 0$ holds $\forall i \in \{ 1,...,L \}$. Based on the assumptions made and the reverse triangle inequality $||a| - |b|| \leq |a-b|$, we have
\begin{subequations}
	\begin{align}
		&f(r_{i} - d_{i}^{*}) + \tfrac{\rho}{2} {\big( \big\| \bm{v}_i^{(k+1)} \big\|_2 - d_{i}^{*} \big)}^2 \\
		&~= f(|r_{i} - d_{i}^{*}|) + \tfrac{\rho}{2} {\big( \big| \big\| \bm{v}_i^{(k+1)} \big\|_2 - d_{i}^{*} \big| \big)}^2 \\ 
		&~\geq f(||r_{i}| - |d_{i}^{*}||) + \tfrac{\rho}{2} {\big( \big| \big| \big\| \bm{v}_i^{(k+1)} \big\|_2 \big| - \big| d_{i}^{*} \big| \big| \big)}^2 \label{ch:ADMM:eqpps1_ineq_dsubp} \\
		&~= f(|r_{i}| - |d_{i}^{*}|) + \tfrac{\rho}{2} {\big( \big| \big\| \bm{v}_i^{(k+1)} \big\|_2 \big| - \big| d_{i}^{*} \big| \big)}^2 \\
		&~= f(r_{i} - |d_{i}^{*}|) + \tfrac{\rho}{2} {\big( \big\| \bm{v}_i^{(k+1)} \big\|_2 - \big| d_{i}^{*} \big| \big)}^2.
	\end{align}
\end{subequations}
which means that the objective function value associated with the global optimum $\bm{d} = \bm{d}^{*}$ is greater than or equal to its counterpart associated with $\bm{d} = \textup{abs}(\bm{d}^{*})$, where $\textup{abs}(\cdot)$ stands for the element-wise absolute value function. Namely, $\bm{d}^{*}$ will no longer be the globally optimal solution if the inequality in (\ref{ch:ADMM:eqpps1_ineq_dsubp}) holds strictly. It then follows that $\geq$ in (\ref{ch:ADMM:eqpps1_ineq_dsubp}) should degrade into $=$, which holds if and only if $d_{i}^{*} = |d_{i}^{*}|, \forall i \in \{ 1,...,L \}$ since the sum of two strictly monotonic functions of the same kind of monotonicity is still a monotonic function. Therefore, $d_{i}^{*} \geq 0$ is tacitly satisfied $\forall i \in \{ 1,...,L \}$. $\square$

\section{Proof of Proposition 2}
\label{Sect_AppenB}
Let $(\bm{x}^{*},\bm{d}^{*},\bm{\beta}^{*})$ be the point corresponding to the globally optimal solution to (\ref{ch:ADMM:eqell_p_ref_2}). To prove the proposition, it suffices to show that $d_{i}^{*} \geq 0$ holds $\forall i \in \{ 1,...,L \}$. Based on the assumptions made and the reverse triangle inequality, we have
\begin{subequations}
	\begin{align}
		&\sum_{i=1}^{L}f(r_{i} - d_{i}^{*}) = \sum_{i=1}^{L}f(|r_{i} - d_{i}^{*}|) \\
		&~\geq \sum_{i=1}^{L}f(||r_{i}| - |d_{i}^{*}||) \label{ch:ADMM:eqpps1_ineq} \\
		&~= \sum_{i=1}^{L}f(|r_{i}| - |d_{i}^{*}|) = \sum_{i=1}^{L}f(r_{i} - |d_{i}^{*}|).
	\end{align}
\end{subequations}
This implies that the objective function value associated with the global optimum $(\bm{x}^{*},\bm{d}^{*},\bm{\beta}^{*})$ is greater than or equal to that achieved by the feasible solution point $\big( \bm{x}^{*},\textup{abs}(\bm{d}^{*}),\hat{\bm{\beta}}^{*} \big)$, where $\hat{\bm{\beta}} = \big[ \hat{\bm{\beta}}_{1}^T,...,\hat{\bm{\beta}}_{L}^T \big]^T \in \mathbb{R}^{HL}$, and $\hat{\bm{\beta}}_{i}^{*} = {\bm{\beta}}_{i}^{*} \cdot \textup{sgn}(d_{i}^{*})$ for $i=1,...,L$. In other words, $(\bm{x}^{*},\bm{d}^{*},\bm{\beta}^{*})$ will not be the globally optimal solution point anymore if the inequality in (\ref{ch:ADMM:eqpps1_ineq}) holds strictly. It then follows that the symbol $\geq$ in (\ref{ch:ADMM:eqpps1_ineq}) should be replaced by $=$, which holds if and only if $d_{i}^{*} = |d_{i}^{*}|, \forall i \in \{ 1,...,L \}$. Therefore, $d_{i}^{*} \geq 0$ is tacitly satisfied $\forall i \in \{ 1,...,L \}$. $\square$

\section{Proof of Theorem 1}
\label{Sect_AppenC}
Since (\ref{ch:ADMM:eqUpd_UV2}) renders the solution to the $\bm{\beta}$-subproblem in each of the ADMM iterations certainly feasible, the condition (\ref{ch:ADMM:eqKKTcon_beta}) is satisfied. On the other hand, based on the dual variable update (\ref{ch:ADMM:eqdlambda}) and our assumption about $\lim_{k \rightarrow \infty} \bm{\lambda}^{(k)}$ in (\ref{ch:ADMM:eqlim_assump}), it is straightforward to deduce that the condition (\ref{ch:ADMM:eqKKTcon_cons}) is satisfied as well. We now proceed to check the remaining conditions (\ref{ch:ADMM:eqKKTcon_Lag1})--(\ref{ch:ADMM:eqKKTcon_Lag3}). As ${\bm{x}}^{(k+1)}$ and $\big( {\bm{d}}^{(k+1)}, \bm{\beta}^{(k+1)} \big)$ are the minimizers of the subproblems (\ref{ch:ADMM:eqpx}) and (\ref{ch:ADMM:eqpdbeta}), respectively, the following relations hold:
\begin{subequations}{\label{ch:ADMM:eqmin_rela}}
	\begin{align}
		&\bm{\nabla}_{\bm{x}} \mathcal{L}_\rho \big(\bm{x}^{(k+1)}, {\bm{d}}^{(k)}, \bm{\beta}^{(k)}, \bm{\lambda}^{(k)} \big) = \bm{0}_{H},\label{ch:ADMM:eqmin_rela_a}\\
		&\bm{\nabla}_{\bm{d}} \mathcal{L}_\rho \big(\bm{x}^{(k+1)}, {\bm{d}}^{(k+1)}, \bm{\beta}^{(k+1)}, \bm{\lambda}^{(k)} \big) = \bm{0}_{L},\label{ch:ADMM:eqmin_rela_b}\\
		&\bm{\nabla}_{\bm{\beta}} \mathcal{L}_\rho \big(\bm{x}^{(k+1)}, {\bm{d}}^{(k+1)}, \bm{\beta}^{(k+1)}, \bm{\lambda}^{(k)} \big) = \bm{0}_{HL}.\label{ch:ADMM:eqmin_rela_c}
	\end{align}
\end{subequations}
Putting (\ref{ch:ADMM:eqlim_assump}), (\ref{ch:ADMM:eqmin_rela_a}), and the verified condition (\ref{ch:ADMM:eqKKTcon_cons}) together, we have
\begin{align}{\label{ch:ADMM:eqverify_equa_1}}
	&\lim_{k \rightarrow \infty} \bm{\nabla}_{\bm{x}} \mathcal{L}_\rho \big(\bm{x}^{(k+1)}, {\bm{d}}^{(k)}, \bm{\beta}^{(k)}, \bm{\lambda}^{(k)} \big) = \bm{0}_H = \lim_{k \rightarrow \infty} \sum_{i=1}^{L} \big[ \bm{\lambda}_{i}^{(k)} + \rho \big( \bm{x}^{(k+1)} - \bm{x}_{i} - \bm{\beta}_{i}^{(k)} \cdot d_{i}^{(k)} \big) \big] \nonumber\\
	&~=\sum_{i=1}^{L} \bm{\lambda}_{i}^{\star} =\bm{\nabla}_{\bm{x}} \hat{\mathcal{L}} (\bm{x}^{\star}, \bm{d}^{\star},\bm{\beta}^{\star}, \bm{\lambda}^{\star}).
\end{align}
Analogously, it follows from (\ref{ch:ADMM:eqlim_assump}), (\ref{ch:ADMM:eqKKTcon_cons}), and (\ref{ch:ADMM:eqmin_rela_b}) that
\begin{align}{\label{ch:ADMM:eqverify_equa_2}}
	&\lim_{k \rightarrow \infty} \bm{\nabla}_{\bm{d}} \mathcal{L}_\rho \big(\bm{x}^{(k+1)}, {\bm{d}}^{(k+1)}, \bm{\beta}^{(k+1)}, \bm{\lambda}^{(k)} \big) = \bm{0}_{L} = \big[ -f^{\prime}(\cdot)-{(\bm{\lambda}_{1}^{\star})}^T\bm{\beta}_{1}^{\star},...,-f^{\prime}(\cdot)-{(\bm{\lambda}_{L}^{\star})}^T\bm{\beta}_{L}^{\star} \big]^T \nonumber\\
	&~=\bm{\nabla}_{\bm{d}} \hat{\mathcal{L}} (\bm{x}^{\star}, \bm{d}^{\star},\bm{\beta}^{\star}, \bm{\lambda}^{\star}),
\end{align}
and (\ref{ch:ADMM:eqlim_assump}), (\ref{ch:ADMM:eqKKTcon_cons}), and (\ref{ch:ADMM:eqmin_rela_c}) that
\begin{align}{\label{ch:ADMM:eqverify_equa_3}}
	&\lim_{k \rightarrow \infty} \bm{\nabla}_{\bm{\beta}} \mathcal{L}_\rho \big(\bm{x}^{(k+1)}, {\bm{d}}^{(k+1)}, \bm{\beta}^{(k+1)}, \bm{\lambda}^{(k)} \big) = \bm{0}_{HL} = \big[ - d_{1}^{\star}(\bm{\lambda}_{1}^{\star})^T,...,- d_{L}^{\star}(\bm{\lambda}_{L}^{\star})^T \big]^T \nonumber\\
	&~=\bm{\nabla}_{\bm{\beta}} \hat{\mathcal{L}} (\bm{x}^{\star}, \bm{d}^{\star},\bm{\beta}^{\star}, \bm{\lambda}^{\star}).
\end{align}
The conditions (\ref{ch:ADMM:eqKKTcon_Lag1})--(\ref{ch:ADMM:eqKKTcon_Lag3}) are verified by the above equalities.

Thus far, we have wrapped up the proof for the core component of Theorem 1. Our attention will now shift towards verifying the linear independence constraint qualification (LICQ), which has been one of the most commonly adopted constraint qualifications and is necessary to guarantee that an optimal solution will indeed adhere to the KKT conditions \cite{GWachsmuth}. Put differently, the significance of this verification lies in the fact that an optimization problem can potentially possess an optimal solution while lacking KKT points \cite{LICQeg}. Moreover, the uniqueness of the Lagrange multipliers can be ensured by the verification of LICQ \cite{JNocedal}. In our situation, LICQ pertains to the linear independence of constraint gradients at the local solution point $\bm{y}^{\star} = \big[ (\bm{x}^{\star})^T,(\bm{d}^{\star})^T,(\bm{\beta}^{\star})^T \big]^T$:
\begin{align}{\label{ch:ADMM:eqGradbmh}}
& \tfrac{\partial (\bm{x}\!-\!\bm{x}_i\!-\!\bm{\beta}_i\!\cdot\!d_{i})}{\partial \bm{y}} \Big|_{\bm{y} = \bm{y}^{\star}}\!=\!\big[ \bm{I}_{H}, \bm{0}_{H \times (i-1)}, -\bm{\beta}_{i}^{\star}, \bm{0}_{H \times (L-i)}, \bm{0}_{H \times (i-1)H}, -d_{i}^{\star} \cdot \bm{I}_{H}, \bm{0}_{H \times (L-i)H} \big],~i = 1,...,L,
\end{align}
where $\bm{0}_{a \times b}$ (resp. $\bm{I}_{a}$) denotes the $a \times b$ zero (resp. $a \times a$ identity) matrix. In a nontrivial source localization setting, the position of the source is different from those of the sensors \cite{WXiong3}. That is, none of the elements of $\{ \bm{\beta}_{i}^{\star} \}$ should be equal to zero and the same goes for $\{ d_{i}^{\star} \}$. Consequently, it is readily apparent that the LICQ holds true in TOA SL scenarios with practical relevance. $\square$

\section{Proof of Theorem 2}
\label{Sect_AppenD}
In order to study the monotonicity properties of the sequence
\begin{equation}
	\big\{ \mathcal{L}_\rho \big(\bm{x}^{(k)}, \bm{d}^{(k)}, \bm{\beta}^{(k)}, \bm{\lambda}^{(k)} \big) \big\}_{k=1,...},
\end{equation}
we decompose the difference in $\mathcal{L}_\rho$ between two successive ADMM iterations as
\begin{subequations}{\label{ch:ADMM:eqDiff_all}}
	\begin{align}
		&\mathcal{L}_\rho \big( \bm{x}^{(k+1)}, \bm{d}^{(k+1)}, \bm{\beta}^{(k+1)}, \bm{\lambda}^{(k+1)} \big) - \mathcal{L}_\rho \big( \bm{x}^{(k)}, \bm{d}^{(k)}, \bm{\beta}^{(k)}, \bm{\lambda}^{(k)} \big) \nonumber\\
		&~~=\mathcal{L}_\rho \big(\bm{x}^{(k+1)}, \bm{d}^{(k)}, \bm{\beta}^{(k)}, \bm{\lambda}^{(k)} \big) - \mathcal{L}_\rho \big(\bm{x}^{(k)}, \bm{d}^{(k)}, \bm{\beta}^{(k)}, \bm{\lambda}^{(k)} \big) \label{ch:ADMM:eqDiff1} \\
		&~~~~+ \mathcal{L}_\rho \big( \bm{x}^{(k+1)}, \bm{d}^{(k)}, \bm{\beta}^{(k+1)}, \bm{\lambda}^{(k)} \big) - \mathcal{L}_\rho \big( \bm{x}^{(k+1)}, \bm{d}^{(k)}, \bm{\beta}^{(k)}, \bm{\lambda}^{(k)} \big) \label{ch:ADMM:eqDiff2beta} \\
		&~~~~+ \mathcal{L}_\rho \big( \bm{x}^{(k+1)}, \bm{d}^{(k+1)}, \bm{\beta}^{(k+1)}, \bm{\lambda}^{(k)} \big) - \mathcal{L}_\rho \big( \bm{x}^{(k+1)}, \bm{d}^{(k)}, \bm{\beta}^{(k+1)}, \bm{\lambda}^{(k)} \big) \label{ch:ADMM:eqDiff2d} \\
		&~~~~+\mathcal{L}_\rho \big(\bm{x}^{(k+1)}, \bm{d}^{(k+1)}, \bm{\beta}^{(k+1)}, \bm{\lambda}^{(k+1)} \big) - \mathcal{L}_\rho \big(\bm{x}^{(k+1)}, \bm{d}^{(k+1)}, \bm{\beta}^{(k+1)}, \bm{\lambda}^{(k)} \big) \label{ch:ADMM:eqDiff3}
	\end{align}
\end{subequations}
and analyze their ranges one by one.

For (\ref{ch:ADMM:eqDiff1}) and (\ref{ch:ADMM:eqDiff2beta}), it follows from the $\bm{x}$- and $\bm{\beta}$-updates of the proposed ADMM that
\begin{align}{\label{ch:ADMM:eqDiff1_elab}}
	\mathcal{L}_\rho \big(\bm{x}^{(k+1)}, \bm{d}^{(k)}, \bm{\beta}^{(k)}, \bm{\lambda}^{(k)} \big) - \mathcal{L}_\rho \big(\bm{x}^{(k)}, \bm{d}^{(k)}, \bm{\beta}^{(k)}, \bm{\lambda}^{(k)} \big) \leq 0
\end{align}
and
\begin{align}{\label{ch:ADMM:eqDiff2beta_elab}}
	\mathcal{L}_\rho \big( \bm{x}^{(k+1)}, \bm{d}^{(k)}, \bm{\beta}^{(k+1)}, \bm{\lambda}^{(k)} \big) - \mathcal{L}_\rho \big( \bm{x}^{(k+1)}, \bm{d}^{(k)}, \bm{\beta}^{(k)}, \bm{\lambda}^{(k)} \big) \leq 0
\end{align}
hold by definition.

On the other hand, additional assumptions about $f(\cdot)$ will be required to facilitate the analysis of (\ref{ch:ADMM:eqDiff2d}) and (\ref{ch:ADMM:eqDiff3}). As long as we assume the convexity of $f(r_{i} - d_{i})$ w.r.t. $d_{i}, \forall i \in \{ 1,...,L \}$, it would then be easily verified that
\begin{align}{\label{ch:ADMM:eqAugLwrtd}}
	&\mathcal{L}_\rho \big( \bm{x}^{(k+1)}, \bm{d}, \bm{\beta}^{(k+1)}, \bm{\lambda}^{(k)} \big) = \sum_{i=1}^{L}f( r_{i} - d_{i}) + \sum_{i = 1}^{L} \big( \bm{\lambda}_i^{(k)} \big)^T \big( \bm{x}^{(k+1)} - \bm{x}_i - \bm{\beta}_i^{(k+1)} \cdot d_{i} \big) \nonumber\\
	&~+ \tfrac{\rho}{2} \sum_{i = 1}^{L} {\big\| \bm{x}^{(k+1)} - \bm{x}_i - \bm{\beta}_i^{(k+1)} \cdot d_{i} \big\|}_2^2
\end{align}
is strongly convex w.r.t. $\bm{d}$ (with parameter $M_{2} > 0$), by examining the positive semidefiniteness of the difference between its Hessian and $M_{2} \cdot \bm{I}_{L}$. Applying the equivalent inequality that characterizes strong convexity \cite{SBoyd1} to the points $\bm{d}^{(k+1)}$ and $\bm{d}^{(k)}$, we have for (\ref{ch:ADMM:eqDiff2d})
\begin{align}{\label{ch:ADMM:eqDiff2d_elab_pre}}
	&\mathcal{L}_\rho \big( \bm{x}^{(k+1)}, \bm{d}^{(k)}, \bm{\beta}^{(k+1)}, \bm{\lambda}^{(k)} \big) \geq \mathcal{L}_\rho \big( \bm{x}^{(k+1)}, \bm{d}^{(k+1)}, \bm{\beta}^{(k+1)}, \bm{\lambda}^{(k)} \big) \nonumber\\
	&~~+ \big( \bm{\nabla}_{\bm{d}} \mathcal{L}_\rho \big( \bm{x}^{(k+1)}, \bm{d}, \bm{\beta}^{(k+1)}, \bm{\lambda}^{(k)} \big) \big|_{\bm{d} = \bm{d}^{(k+1)}} \big)^T \big( \bm{d}^{(k)} - \bm{d}^{(k+1)} \big) + \tfrac{M_{2}}{2} {\big\| \bm{d}^{(k)} - \bm{d}^{(k+1)} \big\|}_2^2 \nonumber\\
	&~= \mathcal{L}_\rho \big(\bm{x}^{(k+1)}, \bm{d}^{(k+1)}, \bm{\beta}^{(k+1)}, \bm{\lambda}^{(k)} \big) + \tfrac{M_{2}}{2} {\big\| \bm{d}^{(k)} - \bm{d}^{(k+1)} \big\|}_2^2,
\end{align}
which subsequently leads to
\begin{align}{\label{ch:ADMM:eqDiff2d_elab}}
	\mathcal{L}_\rho \big(\bm{x}^{(k+1)}, \bm{d}^{(k+1)}, \bm{\beta}^{(k+1)}, \bm{\lambda}^{(k)} \big) - \mathcal{L}_\rho \big(\bm{x}^{(k+1)}, \bm{d}^{(k)}, \bm{\beta}^{(k+1)}, \bm{\lambda}^{(k)} \big) \leq -\tfrac{M_{2}}{2} {\big\| \bm{d}^{(k)} - \bm{d}^{(k+1)} \big\|}_2^2.
\end{align}
The remaining difference of two evaluated augmented Lagrangians, (\ref{ch:ADMM:eqDiff3}), can be re-expressed as
\begin{align}{\label{ch:ADMM:eqDiff3_elab}}
	&\mathcal{L}_\rho \big(\bm{x}^{(k+1)}, \bm{d}^{(k+1)}, \bm{\beta}^{(k+1)}, \bm{\lambda}^{(k+1)} \big) - \mathcal{L}_\rho \big(\bm{x}^{(k+1)}, \bm{d}^{(k+1)}, \bm{\beta}^{(k+1)}, \bm{\lambda}^{(k)} \big) \nonumber\\
	&~~= \sum_{i = 1}^{L}\!\big( \bm{\lambda}_i^{(k+1)} - \bm{\lambda}_i^{(k)} \big)^T\!\big( \bm{x}^{(k+1)} - \bm{x}_i - \bm{\beta}_i^{(k+1)} \cdot d_{i}^{(k+1)} \big) = \tfrac{1}{\rho} \sum_{i = 1}^{L} {\big\| \bm{\lambda}_i^{(k+1)} - \bm{\lambda}_i^{(k)} \big\|}_{2}^2
\end{align}
based on (\ref{ch:ADMM:eqdlambda}). Since our $d_i$-minimization step implies
\begin{align}
	&d_i^{(k+1)} = \arg \min_{d_i} \Big[ \big( \bm{\lambda}_i^{(k)} \big)^T \big( \bm{x}^{(k+1)} - \bm{x}_i - \bm{\beta}_i^{(k+1)} \cdot d_{i} \big) \nonumber\\
	&~+ f \big( r_{i} - d_{i} \big) + \tfrac{\rho}{2} {\big\| \bm{x}^{(k+1)} - \bm{x}_i - \bm{\beta}_i^{(k+1)} \cdot d_{i} \big\|}_2^2 \Big],
\end{align}
taking advantage of the convexity of $f(r_{i} - d_{i})$ w.r.t. $d_{i}, \forall i \in \{ 1,...,L \}$ once again we derive
\begin{align}{\label{ch:ADMM:eqhprime_lmbd1}}
	&-f^{\prime} \big( r_{i} - d_{i}^{(k+1)} \big) - \big( \bm{\lambda}_i^{(k)} \big)^T \bm{\beta}_i^{(k+1)} - \rho \big( \bm{\beta}_i^{(k+1)} \big)^T \big( \bm{x}^{(k+1)} - \bm{x}_i - \bm{\beta}_i^{(k+1)} \cdot d_{i}^{(k+1)} \big) \nonumber\\
	&~~= -f^{\prime} \big( r_{i} - d_{i}^{(k+1)} \big) - \big( \bm{\lambda}_i^{(k+1)} \big)^T \bm{\beta}_i^{(k+1)} = 0
\end{align}
and
\begin{align}{\label{ch:ADMM:eqhprime_lmbd2}}
	-f^{\prime} \big( r_{i} - d_{i}^{(k)} \big) - \big( \bm{\lambda}_i^{(k)} \big)^T \bm{\beta}_i^{(k)} = 0.
\end{align}
Utilizing the fact that $\bm{\beta}_i^{(k)}$ and $\bm{\beta}_i^{(k+1)}$ are both unit vectors and our assumptions about $f(r_{i} - d_{i})$, we arrive at
\begin{align}{\label{ch:ADMM:eqDiff3_elab_2}}
	&\mathcal{L}_\rho \big(\bm{x}^{(k+1)}, \bm{d}^{(k+1)}, \bm{\beta}^{(k+1)}, \bm{\lambda}^{(k+1)} \big) - \mathcal{L}_\rho \big(\bm{x}^{(k+1)}, \bm{d}^{(k+1)}, \bm{\beta}^{(k+1)}, \bm{\lambda}^{(k)} \big) \nonumber\\
	&~~=\tfrac{1}{\rho} \sum_{i=1}^{L} \Big\| \bm{\beta}_i^{(k+1)} \cdot \Big( -f^{\prime} \big( r_{i} - d_{i}^{(k+1)} \big) \Big) - \bm{\beta}_i^{(k)} \cdot \Big( -f^{\prime} \big( r_{i} - d_{i}^{(k)} \big) \Big) \Big\|_2^2 \nonumber\\
	&~~\leq \tfrac{1}{\rho} \sum_{i=1}^{L} \Big| \big|f^{\prime} \big( r_{i} - d_{i}^{(k+1)} \big)\big| + \big|f^{\prime} \big( r_{i} - d_{i}^{(k)} \big)\big| \Big|^2 \nonumber\\
	&~~= \tfrac{1}{\rho} \sum_{i=1}^{L} \Big| \big| \nabla_{d_i}f(r_{i} - d_{i})|_{d_i = d_i^{(k+1)}} \big| + \big| \nabla_{d_i}f(r_{i} - d_{i})|_{d_i = d_i^{(k)}} \big| \Big|^2 \nonumber\\
	&~~\leq \tfrac{\bar{M}^2}{\rho} \sum_{i=1}^{L} {\big| d_i^{(k)} - d_i^{(k+1)} \big|}^2 = \tfrac{\bar{M}^2}{\rho} {\big\| \bm{d}^{(k)} - \bm{d}^{(k+1)} \big\|}_2^2
\end{align}
after putting (\ref{ch:ADMM:eqhprime_lmbd1}) and (\ref{ch:ADMM:eqhprime_lmbd2}) into (\ref{ch:ADMM:eqDiff3_elab}).

Plugging (\ref{ch:ADMM:eqDiff1_elab}), (\ref{ch:ADMM:eqDiff2beta_elab}), (\ref{ch:ADMM:eqDiff2d_elab}) and (\ref{ch:ADMM:eqDiff3_elab_2}) into (\ref{ch:ADMM:eqDiff_all}) yields
\begin{align}{\label{ch:ADMM:eqTheo3_final}}
	&\mathcal{L}_\rho \big( \bm{x}^{(k+1)}, \bm{d}^{(k+1)}, \bm{\beta}^{(k+1)}, \bm{\lambda}^{(k+1)} \big) - \mathcal{L}_\rho \big( \bm{x}^{(k)}, \bm{d}^{(k)}, \bm{\beta}^{(k)}, \bm{\lambda}^{(k)} \big) \nonumber\\
	&~~\leq \left( \tfrac{\bar{M}^2}{\rho}-\tfrac{M_{2}}{2} \right) {\big\| \bm{d}^{(k)} - \bm{d}^{(k+1)} \big\|}_2^2.
\end{align}
To rephrase, $\big\{ \mathcal{L}_\rho \big(\bm{x}^{(k)}, \bm{d}^{(k)}, \bm{\beta}^{(k)}, \bm{\lambda}^{(k)} \big) \big\}_{k=1,...}$ is monotonically nonincreasing for $\rho \geq (2\bar{M}^2)/M_{2}$ under our additional assumptions about the loss function. $\square$

\section{Proof of Theorem 3}
\label{Sect_AppenE}
Combining the convexity of $f(r_{i} - d_{i})$ w.r.t. $d_{i}$ with the $M_{1}$-Lipschitz continuity of its gradient gives \cite[Lemma 4]{XZhou}
\begin{align}{\label{ch:ADMM:eqLB_1}}
	f(r_{i}-\eta_{1}) \leq f(r_{i} - \eta_{2}) - f^{\prime}(r_{i} - \eta_{2})(\eta_{1} - \eta_{2}) + \tfrac{M_{1}}{2} (\eta_{1} - \eta_{2})^2,~\forall \eta_1, \eta_2 \in \mathbb{R}.
\end{align}
Having in mind that $\bm{\beta}_i^{(k)}$ is a unit column vector, we construct
\begin{align}{\label{ch:ADMM:eqLB_2}}
	&f \Big( r_{i} - \big( \bm{\beta}_i^{(k)} \big)^T \big( \bm{x}^{(k)} - \bm{x}_i \big) \Big) \leq f \big( r_{i} - d_{i}^{(k)} \big) - f^{\prime} \big( r_{i} - d_{i}^{(k)} \big) \big( \big( \bm{\beta}_i^{(k)} \big)^T \big( \bm{x}^{(k)} - \bm{x}_i \big) - d_{i}^{(k)} \big) \nonumber\\
	&~+ \tfrac{M_{1}}{2} \Big( \big( \bm{\beta}_i^{(k)} \big)^T \big( \bm{x}^{(k)} - \bm{x}_i \big) - d_{i}^{(k)} \Big)^2
\end{align}
from (\ref{ch:ADMM:eqLB_1}) by letting $\eta_{1}$ and $\eta_{2}$ be $\big( \bm{\beta}_i^{(k)} \big)^T \big( \bm{x}^{(k)} - \bm{x}_i \big)$ and $d_{i}^{(k)}$, respectively. Plugging (\ref{ch:ADMM:eqhprime_lmbd2}) into (\ref{ch:ADMM:eqLB_2}) further deduces
\begin{align}{\label{ch:ADMM:eqLB_3}}
	&f \Big( r_{i} - \big( \bm{\beta}_i^{(k)} \big)^T \big( \bm{x}^{(k)} - \bm{x}_i \big) \Big) \leq f \big( r_{i} - d_{i}^{(k)} \big) + \big( \bm{\lambda}_i^{(k)} \big)^T \big( \bm{x}^{(k)} - \bm{x}_i - \bm{\beta}_i^{(k)} \cdot d_{i}^{(k)} \big) \nonumber\\
	&~~+ \tfrac{M_{1}}{2} \Big( \big( \bm{\beta}_i^{(k)} \big)^T \big( \bm{x}^{(k)} - \bm{x}_i \big) - d_{i}^{(k)} \Big)^2.
\end{align}
Based on (\ref{ch:ADMM:eqLB_3}), we have
\begin{subequations}{\label{ch:ADMM:eqLB_4}}
	\begin{align}
		&\mathcal{L}_\rho \big( \bm{x}^{(k)}, \bm{d}^{(k)}, \bm{\beta}^{(k)}, \bm{\lambda}^{(k)} \big) = \sum_{i=1}^{L}f \big( r_{i} - d_{i}^{(k)} \big) + \sum_{i = 1}^{L} \big( \bm{\lambda}_i^{(k)} \big)^T \big( \bm{x}^{(k)} - \bm{x}_i - \bm{\beta}_i^{(k)} \cdot d_{i}^{(k)} \big) \nonumber\\
		&~~+ \tfrac{\rho}{2} \sum_{i = 1}^{L} {\big\| \bm{x}^{(k)} - \bm{x}_i - \bm{\beta}_i^{(k)} \cdot d_{i}^{(k)} \big\|}_2^2 \\
		&~\geq \sum_{i = 1}^{L} \Big[ \underbrace{f \Big( r_{i} - \big( \bm{\beta}_i^{(k)} \big)^T \big( \bm{x}^{(k)} - \bm{x}_i \big) \Big)}_{\textup{1st part}} + \underbrace{\tfrac{(\rho - M_{1})}{2} {\big\| \bm{x}^{(k)} - \bm{x}_i - \bm{\beta}_i^{(k)} \cdot d_{i}^{(k)} \big\|}_2^2}_{\textup{2nd part}} \Big]. \label{ch:ADMM:eqLB_4_c}
	\end{align}
\end{subequations}
Under the symmetry and monotonicity assumptions earlier made about $f(\cdot)$ in Proposition 1, it is not hard to find that the first part of the equation enclosed within square brackets in (\ref{ch:ADMM:eqLB_4_c}) is bounded from below. The same can also be said of the second part for $\rho \geq M_{1}$ (in fact, $\rho$ should satisfy $\rho \geq \max(M_{1}, (2\bar{M}^2)/M_{2})$ so that Theorem 2 remains valid), trivially, as it is lower bounded by 0. $\square$

\section{Proof of Theorem 4}
\label{Sect_AppenF}
First of all, it follows from (\ref{ch:ADMM:eqTheo3_final}) and Corollary 1 that 
\begin{align}{\label{ch:ADMM:eqSC_1}}
	&\lim_{k \rightarrow \infty} {\big\| \bm{d}^{(k+1)} - \bm{d}^{(k)} \big\|}_2^2 \leq \lim_{k \rightarrow \infty} \left( \tfrac{\bar{M}^2}{\rho}-\tfrac{M_{2}}{2} \right)^{-1} \cdot \big[ \mathcal{L}_\rho \big( \bm{x}^{(k+1)}, \bm{d}^{(k+1)}, \bm{\beta}^{(k+1)}, \bm{\lambda}^{(k+1)} \big) \nonumber\\
	&~~- \mathcal{L}_\rho \big( \bm{x}^{(k)}, \bm{d}^{(k)}, \bm{\beta}^{(k)}, \bm{\lambda}^{(k)} \big) \big] = 0.
\end{align}
Combining (\ref{ch:ADMM:eqDiff3_elab}), (\ref{ch:ADMM:eqDiff3_elab_2}), and (\ref{ch:ADMM:eqSC_1}), we have for $\bm{\lambda}$ 
\begin{align}{\label{ch:ADMM:eqSC_2}}
	\lim_{k \rightarrow \infty} \tfrac{1}{\rho} \sum_{i = 1}^{L} {\big\| \bm{\lambda}_i^{(k+1)} - \bm{\lambda}_i^{(k)} \big\|}_{2}^2 \leq \lim_{k \rightarrow \infty} \tfrac{\bar{M}^2}{\rho} {\big\| \bm{d}^{(k+1)} - \bm{d}^{(k)} \big\|}_2^2 = 0
\end{align}
as well. Akin to (\ref{ch:ADMM:eqAugLwrtd}) and (\ref{ch:ADMM:eqDiff2d_elab_pre}), the strong convexity characterizing inequality (with parameter $M_{3} > 0$) for 
\begin{align}{\label{ch:ADMM:eqSC_3}}
	&\mathcal{L}_\rho \big( \bm{x}, \bm{d}^{(k)}, \bm{\beta}^{(k)}, \bm{\lambda}^{(k)} \big) = \sum_{i=1}^{L}f \big( r_{i} - d_{i}^{(k)} \big) + \sum_{i = 1}^{L} \big( \bm{\lambda}_i^{(k)} \big)^T \big( \bm{x} - \bm{x}_i - \bm{\beta}_i^{(k)} \cdot d_{i}^{(k)} \big) \nonumber\\
	&~+ \tfrac{\rho}{2} \sum_{i = 1}^{L} {\big\| \bm{x} - \bm{x}_i - \bm{\beta}_i^{(k)} \cdot d_{i}^{(k)} \big\|}_2^2
\end{align}
associated with the points $\bm{x} = \bm{x}^{(k+1)}$ and $\bm{x} = \bm{x}^{(k)}$ gives
\begin{align}{\label{ch:ADMM:eqSC_4}}
	&\mathcal{L}_\rho \big( \bm{x}^{(k)}, \bm{d}^{(k)}, \bm{\beta}^{(k)}, \bm{\lambda}^{(k)} \big) \geq \mathcal{L}_\rho \big( \bm{x}^{(k+1)}, \bm{d}^{(k)}, \bm{\beta}^{(k)}, \bm{\lambda}^{(k)} \big) \nonumber\\
	&~~+ \big(\bm{\nabla}_{\bm{x}} \mathcal{L}_\rho \big( \bm{x}, \bm{d}^{(k)}, \bm{\beta}^{(k)}, \bm{\lambda}^{(k)}\!\big) \big|_{\bm{x} = \bm{x}^{(k+1)}} \big)^T \big( \bm{x}^{(k)} - \bm{x}^{(k+1)} \big) + \tfrac{M_{3}}{2} {\big\| \bm{x}^{(k)} - \bm{x}^{(k+1)} \big\|}_2^2 \nonumber\\
	&~= \mathcal{L}_\rho \big(\bm{x}^{(k+1)}, \bm{d}^{(k)}, \bm{\beta}^{(k)}, \bm{\lambda}^{(k)} \big) + \tfrac{M_{3}}{2} {\big\| \bm{x}^{(k)} - \bm{x}^{(k+1)} \big\|}_2^2,
\end{align}
which implies
\begin{align}{\label{ch:ADMM:eqSC_5}}
	\lim_{k \rightarrow \infty} {\big\| \bm{x}^{(k+1)} - \bm{x}^{(k)} \big\|}_2^2 = 0
\end{align}
by invoking Corollary 1 once more. Likewise, we can show
\begin{align}{\label{ch:ADMM:eqSC_6}}
	\mathcal{L}_\rho \big( \bm{x}^{(k+1)}, \bm{d}^{(k)}, \bm{\beta}^{(k)}, \bm{\lambda}^{(k)} \big) \geq \mathcal{L}_\rho \big( \bm{x}^{(k+1)}, \bm{d}^{(k)}, \bm{\beta}^{(k+1)}, \bm{\lambda}^{(k)} \big) + \tfrac{M_{4}}{2} {\big\| \bm{\beta}^{(k)} - \bm{\beta}^{(k+1)} \big\|}_2^2,
\end{align}
and
\begin{align}{\label{ch:ADMM:eqSC_7}}
	\lim_{k \rightarrow \infty} {\big\| \bm{\beta}^{(k+1)} - \bm{\beta}^{(k)} \big\|}_2^2 = 0
\end{align}
for some parameter $M_{4} > 0$, following a procedure similar to (\ref{ch:ADMM:eqSC_3})--(\ref{ch:ADMM:eqSC_5}). $\square$

\section*{Acknowledgment}
This work was supported by the state graduate funding coordinated by the University of Freiburg. The first author would like to thank Prof. Moritz Diehl for the inspiring discussions related to the use of ADMM. All three authors wish to express their gratitude to the co-editor-in-chief for his efforts to ensure a rigorous review process and the anonymous reviewers whose insightful feedback greatly improved this work. Special thanks go to Deutsche Bahn for providing necessary onboard power supply that enabled the first author to conduct several additional simulations during his journey to and from Berlin in October 2023, in response to the first-round reviewers' requests.

%\section*{References}

%\bibliography{mybibfile}

\end{document}